\documentclass[iop]{emulateapj}
\usepackage{natbib}
\usepackage{rotating}
\usepackage{color}

\shorttitle{The Fueling Diagram}
\shortauthors{Stark et al.}
\begin{document}


\newcommand{\htwo}{H$_{2}$}

\title{The Fueling Diagram: Linking Galaxy Molecular-to-Atomic Gas
  Ratios to Interactions and Accretion}


\author{David V. Stark\altaffilmark{1}, Sheila
  J. Kannappan\altaffilmark{1}, Lisa H. Wei\altaffilmark{2}, Andrew
  J. Baker\altaffilmark{3}, Adam K. Leroy\altaffilmark{4}, Kathleen
  D. Eckert\altaffilmark{1}, and Stuart N. Vogel\altaffilmark{5}}

\altaffiltext{1}{Physics and Astronomy Department, University of North Carolina,
    Chapel Hill, NC 27516}
\altaffiltext{2}{Atmospheric and Environmental Research, 131 Hartwell Avenue, Lexington, MA 02421}
\altaffiltext{3}{Department of Physics and Astronomy, Rutgers, the State University of New Jersey, 136 Frelinghuysen Road, Piscataway, NJ 08854}
\altaffiltext{4}{National Radio Astronomy Observatory, 520 Edgemont Road, Charlottesville, VA 22903}
\altaffiltext{5}{Department of Astronomy, University of Maryland, College Park, MD 20742}




\begin{abstract}

To assess how external factors such as local interactions and fresh
gas accretion influence the global ISM of galaxies, we analyze the
relationship between recent enhancements of central star formation and
total molecular-to-atomic (\htwo/HI) gas ratios, using a
 broad 
sample of field galaxies spanning early-to-late type
morphologies, stellar masses of $10^{7.2}-10^{11.2}M_{\odot}$, and
diverse stages of evolution.  We find that galaxies occupy several
loci in a ``fueling diagram'' that plots \htwo/HI ratio
vs. mass-corrected blue-centeredness, a metric tracing the degree to
which galaxies have bluer centers than the average galaxy at their
stellar mass.  Spiral galaxies of all stellar masses show a positive
correlation between \htwo/HI ratio and mass-corrected
blue-centeredness.  When combined with previous results linking
mass-corrected blue-centeredness to external perturbations, this
correlation suggests a systematic link between local galaxy
interactions and molecular gas inflow/replenishment.  Intriguingly,
E/S0 galaxies show a more complex picture: some follow the same
correlation, some are quenched, and a distinct population of
blue-sequence E/S0 galaxies (with masses below key scales associated
with transitions in gas richness) defines a separate loop in the
fueling diagram.  This population appears to be composed of low-mass
merger remnants currently in late- or post-starburst states, in which
the burst first consumes the H$_2$ while the galaxy center keeps
getting bluer, then exhausts the H$_2$, at which point the burst
population reddens as it ages.  Multiple lines of evidence suggest
connected evolutionary sequences in the fueling diagram. In
particular, tracking {\it total} gas-to-stellar mass ratios within the
fueling diagram provides evidence of fresh gas accretion onto low-mass
E/S0s emerging from their central starburst episodes.  Drawing on a
comprehensive literature search, we suggest that virtually all
galaxies follow the same evolutionary patterns found in our
broad sample.

\end{abstract}





\section{Introduction}
\label{sec:intro}
It has been well documented that stars form in molecular gas
\citep[e.g.,][]{Bigiel08}.  Therefore, understanding what drives the
conversion of hydrogen between its atomic and molecular forms is key
to understanding how galaxies evolve.  The molecular-to-atomic gas
mass ratio (\htwo/HI) varies widely between galaxies, and several
studies have aimed to determine what properties of galaxies -- or
their environments -- play the largest role in the evolution of
\htwo/HI.

Work from the last few decades has revealed correlations between {\it
  global} \htwo/HI and such properties as luminosity (or stellar
mass), total gas mass, morphology, and specific star formation rate
\citep{Kenney89,Thronson89,Young89,Braine93,Casoli98,
  Boselli02,Obreschkow09}.  Unfortunately, these studies have been
largely focused on only certain types of galaxies (e.g., massive
spirals) and/or star-forming/FIR-bright galaxies in the nearby
universe, which are not representative of the galaxy population as a
whole.  More recently, the CO Legacy Database for the GALEX Arecibo
SDSS Survey (COLD GASS) has measured CO emission for a randomly selected
sample of $\sim$300 galaxies with stellar masses
$M_*>10^{10}M_{\odot}$.  COLD GASS finds correlations between \htwo/HI
and structural properties like stellar mass, global stellar mass
surface density, $r$-band light concentration index, and $NUV-r$
color, while also finding that there are thresholds in galaxy
concentration index and global stellar surface mass density above
which detections of molecular gas begin to disappear
\citep{Saintonge11,Kauffmann12}.  While the survey's selection improves upon
past work by focusing on more ``normal'' galaxies, a caveat is that it
only includes massive galaxies.  Moreover, multiple physical
mechanisms may underlie the observed correlations.

The question remains as to what balance of internal and external
processes regulates \htwo/HI in galaxies.  A number of authors have
aimed to explain \htwo/HI completely in terms of local physics within
galaxies, set by their structure and dynamics.  Detailed studies of
atomic and molecular gas in nearby galaxies on kpc scales
\citep{Regan01,Kuno07,Walter08,Leroy09} have shown that \htwo/HI is
correlated with local {\it internal} variables such as galactocentric
radius, stellar and gas surface mass density, radially varying
velocity dispersion and rotation velocity, or combinations of these
\citep{Wong02, Blitz04,Leroy08, Lucero09,Schruba11}.  Dwarf galaxies often show
deviations from these relations, although such deviations are commonly
assumed to reflect the metallicity dependence of the CO-to-H$_2$
conversion factor, X$_{\rm CO}$
\citep{Wilson95,Arimoto96,Bolatto08,Obreschkow09,Wolfire10,Maclow11,Leroy11}. Correlations
of \htwo/HI with internal properties can be used to support models
where the molecular fraction of gas is governed by the equilibrium
between molecule formation on dust grains and destruction by the FUV
background \citep{Elmegreen93, Blitz06, Krumholz09}.  Alternative
theories argue that molecular cloud formation is a very
non-equilibrium process, spurred by gravitational instabilities,
converging gas flows, and/or cloud-cloud collisions
\citep{Tan00,Maclow04,Heitsch08,Pelupessy09}, and that newly formed
molecular clouds fail to reach equilibrium with their environments
before they are destroyed by star formation \citep{Maclow12}.  In this
picture, molecular cloud formation may be highly dependent on drivers
that disrupt gas equilibrium.

Consideration of global dynamical states suggests that external
factors lead to increases in H$_2$.  At the very least,
external perturbations and/or bars can transport gas to the central
regions of galaxies where it can exist at higher densities that
promote the conversion of HI to ${\rm H_2}$ and then stars.  Barred
galaxies, which may be linked to interactions \citep{Gerin90,Miwa98},
display central gas concentrations indicative of inflows
\citep{Sakamoto99,Sheth05}.  Interacting systems display deviations
from typical \htwo/HI and star formation rate relations, with higher
average \htwo/HI than non-interacting galaxies \citep{Kenney89,
  Braine93, Lisenfeld11} and higher star formation rate density than
predicted using the Kennicutt-Schmidt relation defined by normal
spirals \citep{Kennicutt98,Bigiel08}.  Moreover, \citet{Blitz06} find
higher \htwo/HI ratios at a given mid-plane pressure for interacting
systems compared to non-interacting systems.  {\it Post}-starburst
galaxies similarly show excess star formation given their gas content,
likely due to H$_2$ heating and depletion by the starburst
\citep{Leroy06,Robertson08,Wei10b}.  \citet{Papadopoulos10} recreate
such deviations in simulations, linking them to quickly evolving
systems, particularly those that are gas rich and have experienced
recent minor mergers or fresh gas infall.  It should also be noted
that X$_{\rm CO}$ may be lower for centrally concentrated molecular
gas, possibly making the increase in molecular gas due to an
interaction appear even higher \citep{Downes93,Garcia93,Regan00}.

The correlation of galaxy interactions with molecular gas enhancement
(and/or higher CO luminosity per unit molecular gas mass) is an
established result, but an open question remains whether
interaction-induced enhancements are occasional serendipitous events
or the dominant driver of observed \htwo/HI ratios in galaxies.
\citet{KJB} argue that galaxy interactions account for the majority of
recent central star formation enhancements that produce blue central
color excesses, based on the correlation of blue-centered galaxies and
signs of minor mergers/encounters, a result that is confirmed by
\citet{Kauffmann11}.  This correlation implies that blue-centered
galaxies experience gas inflows to feed these star-forming events, a
scenario additionally supported by such galaxies' transient decreased
central metallicities \citep{Kewley06,Kewley10,Rupke10}.  \citet{KJB}
find that roughly 10\% of star-forming galaxies show blue-centered
color gradients, and most have likely experienced a blue-centered
phase at least once in their lifetimes; this percentage increases with
decreasing luminosity.  Additionally, when the systematic trend
towards central reddening at higher luminosity is subtracted (yielding
luminosity-corrected blue-centeredness), the number of galaxies
identified as having enhanced central blueness at fixed luminosity
increases, and the correlation with interactions is stronger.  These
results are interesting when combined with recent analyses of the
rates of minor mergers and flyby encounters, which suggest that
intermediate-to-high mass galaxies experience such frequent
interactions that they can rarely be considered truly isolated
\citep{Maller06,Fakhouri08,Jogee09,Hopkins10,Lotz11,Sinha12}.
Therefore, interaction-induced inflows may play a key role in
molecular gas replenishment for much of the galaxy population.

To explore this idea, this paper examines the relationship between
global \htwo/HI and recent central star formation enhancements
parametrized by mass-corrected blue-centeredness (precisely defined in
\S~\ref{sec:bc}) for a representative sample of galaxies spanning a
wide variety of stellar masses, morphologies, and evolutionary states.
We find a striking relationship between mass-corrected
blue-centeredness and \htwo/HI for nearly all spiral galaxies and a
fraction of E/S0 galaxies, implying a systematic link between external
perturbations and global \htwo/HI ratios.  Intriguingly, our data also
reveal that low-mass blue sequence E/S0 galaxies -- i.e., E/S0
galaxies that fall on the blue-sequence in color versus stellar mass
space with spirals \citep{KGB} -- define an evolutionary track offset
from the main relation toward stronger blue-centered color gradients
at a given \htwo/HI, and trends in these galaxies' total
gas-to-stellar mass ratios along this track suggest the likelihood of
fresh gas accretion during a late- to post-starburst phase.  Thus we
find that several evolutionary tracks can be summarized within a
``fueling diagram'' that links the immediate fuel available for star
formation, the total gas, and a metric tracing the recent interactions
that drive the fueling.


\section{Data and Methods}
\label{sec:data}

This section describes our initial sample drawn from the Nearby Field
Galaxy Survey and designed to be representative of intermediate mass
E-Sbc galaxies, followed by the larger literature sample we use to
expand our data set and confirm our results. We also describe our new
CO(1-0) and CO(2-1) observations and our methods for extracting the
useful quantities of gas mass, stellar mass, blue-centeredness, and
mass-corrected blue-centeredness.  Finally, we discuss the cuts
applied to our sample to ensure robust results.

All distances are calculated using heliocentric velocity corrected to
the Local Group frame of reference following the method of \citet{Jansen00b}
and assuming ${\rm H_0=70\,km\,s^{-1}\,Mpc^{-1}}$, except in cases where a
more direct distance indicator was available in NED.

\subsection{Samples} 
\label{sec:samples}

\subsubsection{The Nearby Field Galaxy Survey}
\label{sec:nfgs}

Our primary sample comes from the Nearby Field Galaxy Survey (Jansen
et al. 2000a,b; see also Jansen \& Kannappan 2001), a set of $\sim$200
galaxies selected to span a broad range of $B$-band luminosities and
morphologies.  The data products of the original survey include $UBR$
surface photometry and optical spectroscopy
\citep{Jansen00a,Jansen00b,Kannappan01b,Kannappan02,KGB}.  The sample
also includes extensive supporting data relevant to this study.
Roughly 90\% of galaxies have Sloan Digital Sky Survey (SDSS) DR8
optical imaging \citep{Aihara11}.  All have near infrared (NIR) data
from the 2 Micron All Sky Survey (2MASS; Skrutskie et al. 2006) while
55\% have deeper {\it Spitzer} Infrared Array Camera (IRAC; Fazio et
al. 2004) imaging, mainly from \citet{Sheth10}, \citet{Moffett12} and
\citet[submitted, hereafter K13s]{Kannappan13}.  In addition, all
galaxies have single dish 21cm data (\citealt{Wei10a}, K13s).

For this study, we analyze the 35 out of 39 NFGS galaxies with CO data
that pass the useability cuts applied to our final sample (see
\S~\ref{sec:useability}).  This subset of the NFGS has a stellar mass
range of $10^{8.8}-10^{10.5}\;M_{\odot}$, morphologies ranging from E
to Sbc, and diverse states of star formation (e.g., starbursting,
post-starburst, quiescent).  Most of the CO data for this sample came
from new observations, and unlike many previous investigations of
molecular gas in galaxies, our NFGS CO observations were {\it not} in
general designed to emphasize CO-bright galaxies, but were instead
designed to reach strong, scientifically useful upper limits in the
case of CO non-detections.

\subsubsection{Literature Sample}
\label{sec:lit}
To strengthen our results, we supplement our sample with galaxies from
the literature with available CO, HI, and multi-band imaging data.

Our literature sample includes galaxies from three large surveys: the
Spitzer Near Infrared Galaxy Survey (SINGS; \citealt{Kennicutt03}),
ATLAS-3D \citep{young11}, and COLD GASS \citep{Saintonge11}. These
surveys are dominated by high mass galaxies, so to supplement the low
mass end of our data set, we add galaxies from \citet{Barone00},
\citet{Garland05}, \citet{Leroy05}\footnotemark[6], \citet{Taylor98},
and \citet{KGB}.  Most of these references are themselves the sources
of the CO data, although some CO data for galaxies in SINGS and
\citet{KGB} come from \citet{Leroy09}, \citet{Albrecht07},
\citet{Sage07}, \citet{Zhu99}, or our own observations
  (\S~\ref{sec:h2data}). HI data often come from the same source as
  the CO data, or else from alternate sources in the literature or
  HyperLeda
  \citep{Huchtmeier95,Smoker00,Salzer02,Paturel03,Garland04,Springob05,Meurer06,Walter08,Catinella10,Haynes11,Catinella12,Serra12}.
  All optical data come from the SDSS DR8, except for a subset of the
  SINGS sample outside the SDSS footprint that has $BVRI$ photometry
  \citep{Munoz-Mateos09}.  NIR imaging data is available for all
  galaxies from 2MASS.

Combined, these data sources bring our full sample (only
considering galaxies that have all necessary data) to 627 galaxies.
However, our total decreases to 323 galaxies after we institute a
number of useability cuts (see \S~\ref{sec:useability}).  

\footnotetext[6]{We do not include marginal detections due to the
  authors' predictions of a high false positive rate.}

\subsection{New Molecular Gas Data}
\label{sec:h2data}

CO(1-0) data, which we use to estimate molecular gas masses, already
existed for a handful of our NFGS galaxies prior to this study
\citep{Sage92,Wei10b}.  The rest of the NFGS sample was observed with
the Institut de Radioastronomie Millim\'etrique (IRAM) 30m and Arizona
Radio Observatory (ARO) 12m single dish telescopes.  Total integration
times were set by how long it took to reach reasonable integrated
signal-to-noise ratios (S/N$>5$) or strong upper limits (yielding
\htwo/HI$<$0.05) on the CO flux.  Most observations were single
pointings toward galaxy centers, but offset positions were
observed for a handful of larger galaxies.

Initial observing runs on the IRAM 30m took place in Fall 2008 and
used the ABCD receivers to observe the CO(1-0) and CO(2-1) lines at
115 and 230 GHz simultaneously in both polarizations.  The 4 MHz
filter bank provided 1\,GHz bandwidth at velocity resolutions of
$\sim$10.4 km s$^{-1}$ and $\sim$5.2 km s$^{-1}$ at these two
frequencies.  Further observations were taken in Fall/Winter 2009/2010
with the newly commissioned EMIR receiver in conjunction with the
WILMA backend, which supplied 2 MHz resolution and a total bandwidth
of 3.7 GHz.  For all observations, wobbler switching was used with a
throw of 2\arcmin and individual scans of 6 minutes.  The data were
calibrated via observations of an ambient temperature load.  The
absolute calibration is accurate to 15--20\%.  The
half-power beam widths are 22\arcsec and 12\arcsec at 115 GHz and 230
GHz respectively.

The ARO 12m observations took place between December 2010 and April
2011.  We used the ALMA 3mm receiver in conjunction with both 2 MHz
filterbanks (one for each polarization), which provided a total
bandwidth of 512 MHz.  We simultaneously used the Millimeter Auto
Correlator (MAC), with a resolution of 781.2 kHz and a total bandwidth
of 800 MHz.  Observations were carried out in beam switching mode with
typical throws of 2\arcmin--4\arcmin, and individual scans of 6
minutes.  The data were calibrated using measurements of a noise diode
between scans, and galaxies with previous observations were used to
check the calibration.  Some of our ARO time was used to obtain CO
data for five extra galaxies in our literature sample.

The data were reduced using CLASS\footnotemark[7].  Scans were
averaged together and any bad channels flagged.  The spectra were
Hanning smoothed to a resolution of 10.4 km\,s$^{-1}$.  Baselines were
then fit to emission free regions of the spectrum, with polynomials of
order $<$5.

Integrated fluxes and other measured quantities are reported in Tables
1 and 2.  We convert the IRAM 30m data from the measured T$_{\rm A}^*$ scale
to Janskys using Jy/K = 6.12 and 6.3 for the EMIR and ABCD receivers
measured at 115 GHz, and Jy/K = 7.86 and 7.5 for the EMIR and ABCD
receivers measured at 230 GHz (see IRAM 30m
documentation\footnotemark[8]).  The ARO data are initially recorded
in the T$_{\rm R}^*$ scale, which we then convert to Janskys using Jy/K =
38.3 (see the ARO 12m documentation\footnotemark[9]).

\footnotetext[7]{{\it http://www.iram.fr/IRAMFR/GILDAS}}
\footnotetext[8]{{\it http://www.iram.es/IRAMES/mainWiki/Iram30mEfficiencies}}
\footnotetext[9]{{\it http://aro.as.arizona.edu/12\_obs\_manual/chapter\_3.htm\#3.\_Receivers\_ }}

Fluxes were determined by summing the channels within the line
profile.  The specific integration ranges are given in Tables 1 and 2
and were judged by eye for each case.  If the profile edges were
unclear, we made the velocity ranges large enough to ensure all flux
was included without any doubt and also yield a more conservative
error estimate.  The uncertainty in the total flux measurement is
given by

\begin{equation}
\sigma_{f}=\sigma_{rms} \Delta V\sqrt{N_{chan}}
\end{equation}
Here, $\sigma_{rms}$ is the rms noise of the spectrum in Jy as
measured from line-free channels, $\Delta V$ is the velocity
resolution in km s$^{-1}$, and $N_{chan}$ is the number of channels in
the integration.  For non-detections, we take upper limits to be
3$\sigma_{f}$, measured over a velocity range defined by the larger of
the HI line width or an equivalent linewidth from an H$\alpha$
or stellar rotation curve (see K13s).

CO linewidths (W$_{50}$) are determined by finding where the data are
greater than 0.5 times the peak flux minus the rms noise for 3
consecutive channels, starting from the outside of the emission line
and working inwards.  The final left and right edges for width
determination are linear interpolations to get the fractional channels
where the data cross the line height.  Heliocentric velocities are
defined to be midway between the two edges found by the above
algorithm.  Following the examples of \citet{Schneider86} and
\citet{Fouque90}, we estimate line width uncertainties by generating a
series of artificial observations over a model grid with varying line
steepness and peak signal-to-noise ratios.  At our resolution of 10.4
km\,s$^{-1}$, the standard deviation of the linewidth measurements at
each grid point is approximately described by
\begin{equation}
\sigma_{W_{50}}=6.8\frac{P^{0.75}}{\left(\rm S/N\right)^{0.9}}
\end{equation}
where $P$ is the steepness parameter (defined as
$P=\left(W_{50}-W_{20}\right)/2$) and S/N is the peak signal-to-noise
ratio.  We stress that linewidths for profiles with peak S/N$<6$
become extremely unreliable, and should be used with caution.  We
include linewidth measurements in tables 1 and 2 for completeness, but
these measurements never enter into the analysis of this paper.

\subsection{Methods}
\label{sec:methods}

\subsubsection{Optical and NIR Photometry}
\label{sec:photometry}

Optical/NIR photometry is needed to estimate stellar masses and track
recent central star formation enhancements. For our analysis, we do
not use products of the SDSS pipeline since it is prone to shredding
extended sources and does not make use of the most recent background
subtraction algorithm \citep{Blanton11}.  We instead recalculate total
magnitudes using our own custom pipeline, described in greater detail
by Eckert et al.\ (in prep.).  After the downloaded images are
co-added, bright stars and interloping galaxies are masked.  The
masking process is automated with the aid of SExtractor
\citep{Bertin96}, but each mask is inspected by eye and adjusted when
there is clear over- or under-masking using the automatic routine.
The ELLIPSE task in IRAF is used to extract surface brightness
profiles of constant center, PA, and ellipticity, and to sum up the
flux within each isophote.  For the NFGS sample, we adopt the same PAs
and ellipticities used for the $UBR$ photometry \citep{Jansen00b}.
For our literature sample, we adopt the method of Eckert et al.\ (in
prep.), who use ELLIPSE to determine the best PA and ellipticity from
the low surface brightness outer disk of each galaxy using the coadded
$gri$ images.  Models of each galaxy are created with the resulting
surface brightness profiles, and are used to fill in masked regions to
correct the total flux.

Total magnitudes are calculated two ways.  First, we adopt a {\it
  curve of growth} technique very similar to the one outlined in
\citet{Munoz-Mateos09}, where the outer disk values of the enclosed
magnitude and its radial gradient are fit with a linear function.  The
$y$-intercept of this line (i.e., where the enclosed magnitude is no
longer increasing) is the total magnitude.  Total magnitudes are also
calculated by fitting the outer disk of the surface brightness profile
with an exponential function, similar to the method in
\citet{Jansen00b}, except that outlier points are rejected from the
fit.  The total flux is summed up to the last isophote used in the
fit, after which the fit itself is used to estimate the remaining outer
flux.

To estimate systematic uncertainties in our total magnitudes, we
perform each of these total magnitude extrapolations using slightly
differently defined fit ranges (between 1 and 8 times the sky noise,
between 3 and 10 times the sky noise, within a 1 mag arcsec$^{-2}$
range ending at 1 or 3 times the sky noise, and finally using the last
5 data points above 1.5 times the sky noise) and then average the
results (ignoring $>6\sigma$ outliers) to obtain our
final magnitude.  We take the difference of the maximum and minimum
magnitude estimates divided by two (also ignoring outliers) as our
official systematic uncertainty.  This uncertainty is added in
quadrature with the Poisson statistical uncertainty.

The ellipses used for total magnitude calculation best match the shape
of the far outer disk or halo, and are therefore not ideal for
calculating galaxy inclinations.  To estimate photometric
inclinations, ELLIPSE is run a second time where ellipticity is
allowed to vary while PA is kept fixed to the previously used value.
The ellipticity at a surface brightness of 22.5 mag arcsec$^{-2}$ in
the coadded $gri$ images usually reliably traces the
higher surface brightness inner disk and provides accurate
inclinations, which are estimated using
\begin{equation}
\cos{i}=\sqrt{\frac{(b/a)^2-q^2}{1-q^2}}
\end{equation}
Here, $q$ is the intrinsic disk thickness (assumed to be $q = 0.2$),
and $b/a$ is the minor-to-major axis ratio derived from the
ellipticity.  For the NFGS sample, we use the same
inclinations as K13s for consistency, which are based on the same
equation, but not using the b/a measured from SDSS photometry.

$JHK$ magnitudes are also recalculated using the 2MASS imaging data.
Here, we redo the background subtraction for the 2MASS images using a
method similar to that used in the original 2MASS pipeline
\citep{Jarrett00}, where the sky was fit by 3rd order polynomials.
However, we fit the background of each relevant frame only in the
region local to the galaxy of interest (within $\sim5\times R_{25}$),
with the galaxy itself and any stars or background galaxies masked.
We impose the parameters from our first set of optical surface
brightness profiles (center, PA, and ellipticity) to extract NIR
surface brightness profiles.

The two methods for calculating total optical magnitudes described
above are again used to calculate the total NIR magnitudes.  However,
tests comparing the output of these two methods against magnitudes
calculated from much deeper $Spitzer$ IRAC 3.6$\mu$m imaging reveal
that the exponential fit method performs systematically worse than the
curve of growth method when applied to the relatively shallow 2MASS
data.  The fits also perform the best when using a fit range between 3
and 10 times the sky noise.  We thus solely use the curve-of-growth
derived total magnitudes determined using this fit range as our final
estimates, although we still rely on the difference between the
curve-of-growth and exponential fit derived magnitudes to estimate the
systematic error for most galaxies.  For any galaxy for which our
different magnitude estimation techniques yield very different results
(disagreement of more than 0.5 mags), we resort to taking an aperture
magnitude of the galaxy, and then infer the total magnitude by
multiplying by the total-to-aperture flux ratio of a higher S/N
passband (another NIR band like $J$ if possible, otherwise $i$ band).

All optical and NIR magnitudes are corrected for foreground extinction
using the dust maps of \citet{Schlegel98} and the extinction curve of
\citet{Odonnell94}.

\subsubsection{Stellar Masses}
\label{sec:mstars}

Stellar masses are calculated with an improved version of
the method described by \citet{KGB}.  Taking as inputs a combination
of $UBR$, $ugriz$, $JHK$, and 3.6$\mu$m photometry, along with global
optical spectra (or any subset of these inputs that are available; see
K13s and references therein), the stellar mass estimation code fits
mixed young + old stellar populations built from pairs of simple
stellar population (SSP) models from \citet{Bruzual03} assuming a
Salpeter IMF.  Output stellar masses are scaled by 0.7 to match a
``diet'' Salpeter IMF containing fewer low mass stars \citep{Bell03}.
 The only significant change compared to \citet{KGB} is
  the addition of a very young SSP with age 5~Myr to the suite of
  models.  Note also that model SSP pairs can include a
  ``middle-aged'' young SSP, as long as its age is younger than the
  old SSP, which was also true for the \citet{KGB} model grid. In
addition, the $UBR$ zero points have been adjusted for consistency
(see K13s).  The final stellar mass estimate is defined by the median
and 68\% confidence interval of the likelihood weighted mass
distribution over the full model grid.

\begin{figure}
\epsscale{1.2}
\plotone{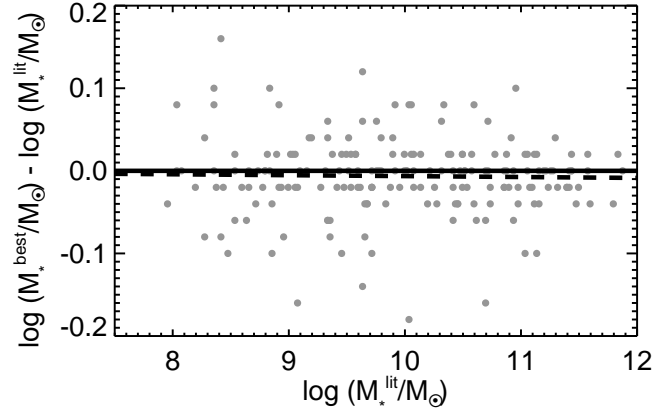}
\caption{Difference between the highest quality stellar masses for the
  NFGS (log M$_*^{\rm best}$, calculated from $UBR$, SDSS, 2MASS, and
  IRAC photometry plus integrated spectroscopy) and stellar masses for
  the same galaxies estimated with data available for our literature
  sample (Log M$_*^{\rm lit}$, calculated using only our custom SDSS
  and 2MASS photometry), testing the quality of stellar masses for our
  literature sample.  The solid horizontal line represents 1:1
  agreement, while the dashed line shows the ordinary least-squares
  fit with outlier rejection, which is not different at a
  statistically significant level.  The scatter has negligible effect
  on our results.}
\label{fig:comparemstars}
\end{figure}

Stellar masses for our literature sample are calculated using only
SDSS and 2MASS photometry.  To determine whether the lack of
spectroscopy leads to any systematic differences in our stellar mass
estimates, we compare the high quality stellar masses from the NFGS
against a second set calculated using only our custom SDSS and 2MASS
photometry.  As shown in Figure~\ref{fig:comparemstars},
  the two methods of estimation are in good agreement.  There is no
  statistical significance in the weak linear trend between the two
  mass estimates as function of stellar mass, and the 1$
\sigma$ scatter between
  the two estimates is 0.04 dex, much less than the typical
  uncertainty of $\sim$0.2 dex for stellar masses.

\subsubsection{Blue-Centeredness}
\label{sec:bc}

To track enhancements in recent central star formation, we rely on a
simple measure of the color gradient referred to as {\it
  blue-centeredness} ($\Delta C$), defined as the outer disk color
from the half-light radius ($r_{50}$) to the 75\% light radius
($r_{75}$) minus the color from the center to $r_{50}$
\citep{Jansen00b}.\footnotemark[10] For future reference, general
discussion of blue-centeredness will use the notation, $\Delta C$,
while specific discussion involving a particular color will note that
color explicitly, e.g. $\Delta(g-r)$.  The radii in this definition
reliably separate bulge/disk colors without being sensitive to
variations in bulge-to-disk ratio \citep{KJB}.  The half-light radius
is a natural separator of inner/outer galaxy growth; for example,
\citet{Peletier96} note shifts in colors and ages at approximately
this radius.  We also stress that this simple measure appears robust
to variations in the dividing radii.  For example, using the 40\% and
90\% enclosed light radii returns approximately the same results.

\footnotetext[10]{Whenever we have reprocessed SDSS images, blue-centeredness is calculated using the second set of ellipse fits, i.e., the ones that are used to estimate inclinations, since they best track the star-forming inner disks (see \S~\ref{sec:photometry}).}

\begin{figure}
\epsscale{1.15}
\plotone{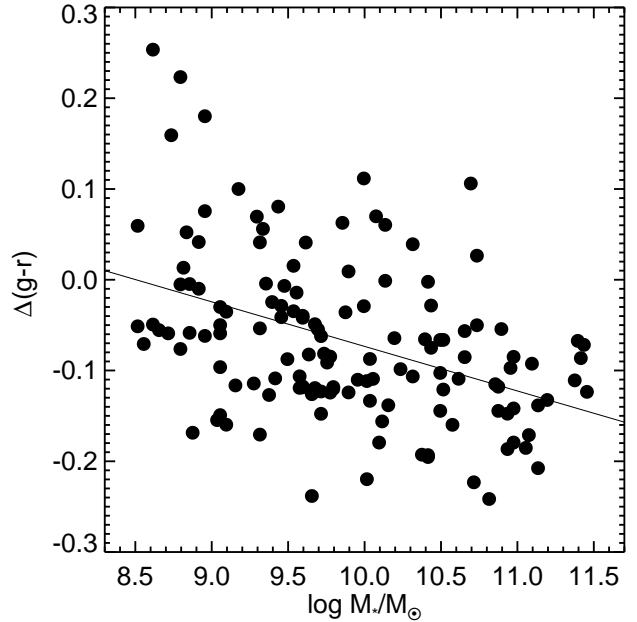}
\caption{$\Delta(g-r)$ versus ${\rm log\,(M_*/M_{\odot})}$,
  illustrating the correlation between blue-centeredness ($\Delta C$)
  and stellar mass. To isolate the effect of recent central star
  formation enhancements, we remove the stellar mass trend
   using an ordinary least-squares fit, minimizing scatter
    in blue-centeredness relative to stellar mass}.  The residuals of
  this fit give {\it mass-corrected blue-centeredness} ($\Delta C^m$),
  which measures outer minus inner color relative to the typical value
  for a galaxy at a given stellar mass, allowing us to trace recent
  central star formation enhancements above the norm.
\label{fig:bcmasscorr}
\end{figure}

Before blue-centeredness is used to measure enhancements in recent
central star formation, we account for other galaxy properties that
may affect the color gradients.  \citet{KJB} find that
blue-centeredness correlates with galaxy luminosity, likely due to the
fact that both metallicity and stellar population age can influence
galaxy color gradients, reddening galaxy centers with increasing
luminosity.  These authors remove the luminosity trend by subtracting
off the fitted relation between $\Delta(B-R)$ and luminosity.   We
follow the same approach, performing an ordinary least-squares fit
\citep{Isobe90} of $\Delta C$ against stellar mass
(Figure~\ref{fig:bcmasscorr}).  We derive the following relations:

\begin{eqnarray}
\hspace{-.5cm}
\Delta(u-r)=\left(-0.114\pm0.025\right)\log{M_*}+\left(0.983\pm0.249\right)\\
\hspace{-.5cm}
\Delta(u-g)=\left(-0.064\pm0.017\right)\log{M_*}+\left(0.556\pm0.167\right)\\
\hspace{-.5cm}
\Delta(g-r)=\left(-0.049\pm0.009\right)\log{M_*}+\left(0.417\pm0.095\right)
\end{eqnarray}

The use of an ordinary least-squares fit of $y$ vs. $x$ is crucial,
since in order to remove any trend with stellar mass, we must minimize
the scatter in $\Delta C$ relative to stellar mass.  Following
\citet{KJB}, we calibrate this mass correction using the NFGS sample
(not just our subset with CO data) since it is our most representative
data set and has the most robust stellar mass estimates.  The fit is
limited to star-forming disk galaxies, identified as galaxies with
morphology S0 or later and either detected H$\alpha$ emission
extending beyond the nucleus or $U-K < 4$, which helps exclude
quenched galaxies (\citealt{Kannappan08}, K13s).  We reject galaxies
with known strong AGN and restrict the fit to $M_{*}>10^{8.5}
M_{\odot}$, avoiding the low M$_*$ tail of the NFGS where all
blue-centeredness states may not be evenly represented.  
  In total, 135 NFGS galaxies were used to derive these relations.
The residuals of this fit give {\it mass-corrected blue-centeredness}
(denoted by $\Delta C^m$).  In simple terms, $\Delta C^m$ is a measure
of the color difference between the inner and outer regions of a
galaxy, relative to the typical color difference for galaxies at a
given stellar mass, where higher values imply bluer centers.  More
physically, this parameter tracks recent central star formation
relative to the typical star-forming galaxy at a given stellar mass.

We find that galaxies with and without strong peculiarities (i.e.,
signs of recent interactions) as defined by \citet{KJB} have different
underlying distributions of $\Delta(g-r)^m$ at a 99\% confidence
level, so our definition of mass-corrected blue-centeredness
successfully recreates the correlation found by these authors using
luminosity-corrected blue-centeredness (their Fig.~8).  However, we
stress that while our mass correction is physically motivated, it is
modest, and the distributions of uncorrected $\Delta(g-r)$ for
peculiar/non-peculiar galaxies differ at confidence level comparable
to that for distributions using $\Delta(g-r)^m$.  Our general results
presented in \S~\ref{sec:results} are still found without the
correction in place, as discussed in \S~\ref{sec:mstars2}.

 Within this paper, the total error budget on $\Delta C^m$
  includes contributions from the measurement error of $\Delta C$, as
  well as additional uncertainties due to the error in $r_{50}$ and
  $r_{75}$, the error in the stellar mass, and the uncertainty in the
  fitted relation between $\Delta C$ and stellar mass.  In the
  majority of cases, the Poisson error is dominant.

 We note that optical color gradients have advantages
   over direct star formation rate tracers (e.g., FUV+24$\mu m$) for
   the purposes of our study.  A practical advantage is that the data
   needed to compute color gradients are readily available for most
   galaxies.  In addition, the longer timescales over which optical
   indicators remain sensitive to enhanced star formation are
   more useful for studying the extended evolution of galaxies.  They
   enable us to analyze the stellar populations beyond the timescale
   of the star formation and gas consumption itself, allowing us to
   note the longer term consequences of recent star formation
   episodes.  We defer a detailed analysis of the timescales of these
   optical indicators to future work, but Figure 11 of \citet{KJB}
   shows a simple model wherein the blue-centeredness fades on the
   order of 0.5--2 Gyr after the central star formation episode ends.
   It should be noted that this model represents a single case, and
   the evolution of blue-centeredness may vary depending on the
   duration and size of the burst as well as the composition of the
   preexisting stellar populations in the bulge and disk.
 
 For consistency and since the vast majority of our galaxies have SDSS
 data, we always quote $\Delta C^m$ using SDSS-equivalent colors.  A
 handful of galaxies from our NFGS sample do not have SDSS data, but
 do have $UBR$ photometry.  We derive the following
   conversions between the NFGS $UBR$ and $ugriz$ systems using colors
   measured within the $B$ band 25 ${\rm mag\,arcsec^{-2}}$ isophote
   for 183 NFGS galaxies:

\begin{eqnarray}
g-r=0.62(B-R)-0.18;\sigma=0.04 \\
u-g=1.16(U-B)+1.08; \sigma=0.16 \\
u-r=0.91(U-R)+0.61; \sigma=0.21
\end{eqnarray}

These relations are useful only for the NFGS due to the possibly
different $UBR$ zeropoints used in the NFGS relative to other
photometric studies (see K13s).  The relations have been corrected for
foreground extinction, although as we are measuring color {\it
  differences}, both foreground extinction corrections and
k-corrections cancel.  Where these conversions are used, their
uncertainties are incorporated into the final error on $\Delta C^m$.

\subsubsection{Internal Extinction Effects}
\label{sec:extinction}

\begin{figure*}
\epsscale{1}
\plotone{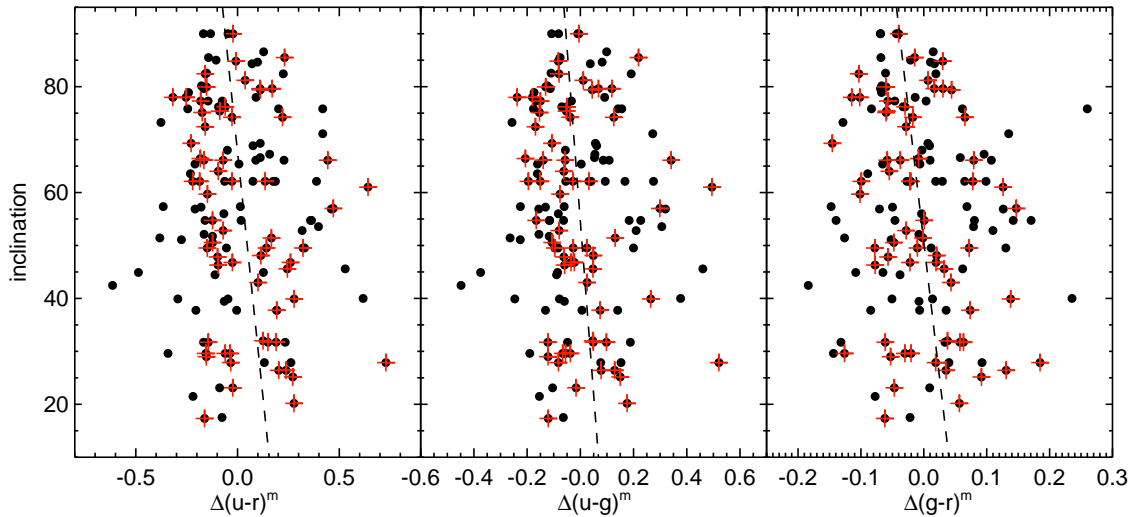}
\caption{Inclination versus $\Delta C^m$ for star-forming disk
  galaxies with i$>$15$^{\circ}$ from the NFGS.  The full sample shows
  no correlation between these two variables, implying no systematic
  effect of inclination (and associated dust extinction) on our
  results. Red crosses mark galaxies with stellar masses above
  $10^{9.7}\,M_{\odot}$, which show a weak correlations between
  inclination and $\Delta C^m$ (3\%, 5\%, and 4\% chance of being
  random for $\Delta (u-r)$, $\Delta (u-g)$, and $\Delta (g-r)$
  respectively).  The dotted lines show the inverse least-squares fits
  (i.e., minimizing the scatter in the $x$-direction) to just the more
  massive galaxies.  Any systematic effects due to inclination appear
  minor.  However, irregular dust structures, not probed with this
  test, will inevitably alter color gradients (see
  \S~\ref{sec:branches} and \S~\ref{sec:mergeleft}).}
\label{fig:inclinbias}
\end{figure*}

Since we make use of optical color gradients, internal dust extinction
is a concern.  A systematic effect from dust may manifest itself as a
dependence on inclination, since higher inclination galaxies show more
dust along the line of sight.  To determine whether internal
extinction is systematically biasing our results, we run Spearman rank
tests on all star-forming disk galaxies in the NFGS with measured
inclinations $>$15$^{\circ}$ after excluding AGN.   We find
  no significant correlation between inclination and $\Delta C^m$.
  Shown in Figure~\ref{fig:inclinbias}, the same rank test for
  galaxies with stellar masses above $\sim10^{9.7}\,M_{\odot}$ -- roughly
  equivalent to the 120 km s$^{-1}$ rotation velocity threshold above
  which dust lanes become more prominent in galaxies
  (\citealt{Dalcanton04}; see also K13s) -- implies somewhat
  significant correlations (3\%, 5\%, and 4\% chance of being random
  for $\Delta (u-r)^m$, $\Delta (u-g)^m$, and $\Delta (g-r)^m$ respectively).
  Though weak correlations may exist, these trends are much smaller
  than the scatter (Fig.~\ref{fig:inclinbias}).  Thus we conclude that
  systematic internal extinction effects should have a minimal effect
  on our results.

We should note that while we find only small systematic effects due to
inclination, dust within galaxies will inevitably alter our color
gradients if it is present.  There are galaxies in our sample for
which visual inspection has revealed significant dust features.  These
galaxies and their impact on our results are discussed in
\S~\ref{sec:results} and \S~\ref{sec:discussion}.

\subsubsection{Gas Masses}
\label{sec:mgas}

To ensure that our data are as uniform as possible, we recalculate all gas
masses using the same formulae.  The HI mass is calculated as \citep{Haynes84}:

\begin{equation}
M_{\rm HI}=2.36\times10^5D^2F_{21}
\end{equation}
Here, $F_{21}$ is the measured 21cm flux in Jy km s$^{-1}$ and $D$ is the
distance to the galaxy in Mpc.  The molecular gas mass is estimated
as \citep{Sanders91}:
\begin{equation}
M_{\rm H_2}= 1.18\times10^4 \left(\frac{X_{\rm CO}}{3\times10^{20}} \right)D^2 F_{\rm CO}
\end{equation}
Here, F$_{\rm CO}$ is the measured CO flux in Jy km s$^{-1}$.  We
assume a constant X$_{\rm CO}$ of $2\times10^{20}$ cm$^{-2}$ (K km
s$^{-1}$)$^{-1}$ \citep{Strong96,Dame01}; see \S~\ref{sec:xco} for an analysis
of how assuming constant X$_{\rm CO}$ may affect our results.  In
further analysis, we multiply all gas masses by a factor of 1.4 to
account for helium.

\subsubsection{Beam Corrections}
\label{sec:beamcorr}

The majority of our CO flux measurements come from single dish
telescopes with single pointings at galaxy centers.  Unlike 21cm
beams, which are typically several times the size of our galaxies, the
CO beams are often comparable to or smaller than our galaxies.
However, since CO scale lengths are typically smaller than optical
scale lengths, this size mismatch is not always a major issue when
trying to estimate total H$_2$ mass.  By modeling the beam pattern of
the telescope and assuming a CO flux distribution, one can estimate
the additional flux not detected by a single pointing.  We apply beam
corrections to our single-dish CO data following the same method as
\citet{Lisenfeld11}.  The aperture correction is defined as the ratio
of the total-to-observed CO intensity,

\begin{equation}
f = I_{\rm CO,total}/I_{\rm CO,observed}
\end{equation}

We assume that the CO distribution $I_{\rm CO}$ follows an exponential
disk,
\begin{equation}
I_{\rm CO}(r)=I_0e^{-r/h_{\rm CO}}
\end{equation}
where $h_{\rm CO}$ is the scale length of the CO emission.  The total CO intensity is simply the integral of this CO distribution to infinity:

\begin{equation}
I_{\rm CO,total}=\int_0^{\infty}{2\pi I_{0}re^{-r/h_{\rm CO}}dr} = 2\pi I_0 h_{\rm CO}^2
\end{equation}

The observed CO intensity is the integral of the CO distribution convolved with the beam pattern.  Assuming a Gaussian beam with HPBW, $\Theta$, this is written:

\begin{eqnarray}
I_{\rm CO,observed}= 4I_{\rm CO}\int^{\infty}_0{\int_0^{\infty}{\exp{\left(-\frac{\sqrt{x^2+y^2}}{h_{\rm CO}}\right)}}}\nonumber \\
\times\exp\left(-\ln{2}\left[\left(\frac{2x}{\Theta}\right)^2 \nonumber +\left(\frac{2y\cos{i}}{\Theta}\right)^2\right]\right)dxdy
\end{eqnarray}


A major uncertainty in this calculation is the value of $h_{\rm CO}$.
Studies of star-forming late-type galaxies have shown $h_{\rm CO}
\sim(0.2\pm0.05)R_{\rm 25}$, where $R_{\rm 25}$ is the $B$-band 25
${\rm mag\,arcsec^{-2}}$ isophotal radius
\citep{Young95,Leroy08,Schruba11}.  We assume this relation holds for
all late type galaxies in our sample.  Whether the same is true for
E/S0 galaxies is uncertain, as their CO scale lengths have not been as
thoroughly studied in the literature.  We examine the CO profiles of
Wei (2010; see also \citealt{Wei10b}), who mapped the CO(1-0)
distribution in a sample of E/S0 galaxies using the CARMA array.  We
find an average value of $h_{\rm CO} = 0.1 R_{\rm 25}$ with a standard
deviation of 0.05$R_{\rm 25}$.  We adopt this scale length for all
E/S0 galaxies in our sample.  It should be noted that these radial
profiles are primarily for blue-sequence E/S0s, not the more common
red-sequence E/S0s, simply because CO is more often detected in the
former.  The two red-sequence E/S0 galaxies in the \citet{Wei10c}
sample have CO scale lengths close to 0.05$R_{\rm 25}$, but two data
points are not enough to warrant separate definitions of $h_{\rm CO}$
for red- and blue-sequence E/S0s.

 We assume uncertainties in h$_{\rm CO}$ of 0.05$R_{25}$ for both
 spiral and E/S0 galaxies.  The additional uncertainty that propagates
 into the beam-corrected H$_2$ mass depends on the size of the galaxy
 relative to the beam.  Within our final sample (see
 \S~\ref{sec:useability}) additional uncertainties are on the order of
 5\%, though some are as large as 25\%. We also note that for galaxies
 in our sample for which we have both single dish CO fluxes and
 resolved CO maps from \citet{Wei10c}, we adopt the CO fluxes from the
 single dish observations and beam-correct these fluxes using the
 measured scale-lengths from the CO maps.

\subsubsection{Useability Criteria}
\label{sec:useability}

After identifying from our NFGS+literature compilation all galaxies
with the necessary optical/NIR imaging as well as 21cm and CO(1-0)
flux measurements (627 galaxies), we institute a number of usability
criteria that galaxies must pass before they are included in our final
sample.  These are: (1) SDSS $r$-band half-light radii larger than 5''
to ensure blue-centeredness calculations are not compromised by
variations in the PSF between the different SDSS passbands; (2)
minimal beam corrections to detected CO fluxes, so that the corrected
CO fluxes are no larger than 1.5 times the measured fluxes (equivalent
to a change in ${\rm H_2/HI}<0.2$ dex), ensuring the estimated H$_2$
masses are minimally dependent on the assumed model of the CO
distribution; (3) strong upper limits on total H$_2$ masses,
i.e. where CO detections are missing but HI detections are not,
\htwo/HI must be $<0.05$.  We do not remove any galaxies with HI upper
limits from our sample.  With these cuts, our NFGS+literature sample
totals 323 galaxies.  Included in this tally are 11 galaxies we judged
to be highly peculiar/interacting whose derived position angles and
ellipticities carry little meaning, and whose blue-centeredness
calculations are therefore untrustworthy; we consider their properties
in \S~\ref{sec:hole}.  We also include an additional 122 ``quenched''
galaxies (two of which are also counted as highly peculiar), which
pass all useability criteria except for having upper limits in both HI
and H$_2$ mass; these are discussed in \S~\ref{sec:sequences}.  For
our NFGS sample, upper limits from the IRAM 30m telescope are
preferred over the ARO 12m telescope since they are always stronger.
Weak upper limits led to the removal of 4 out of 39 NFGS galaxies with
CO data from our final sample, though 3 of the 4 were upper limits
from prior CARMA observations \citep{Wei10b}.  Where we have both IRAM
30m telescope and ARO 12m telescope detections for our NFGS sample, we
prefer the IRAM 30m data as long as the beam correction is less than
1.5.  Otherwise, we use the ARO 12m data.


 It is important to note that while our sample is diverse,
  it is not statistically representative of the galaxy population
  since our combined data set is subject to whatever biases exist in
  past CO studies.  Nonetheless, our useability criteria, which
enforce minimum and maximum apparent radii for many galaxies in our
sample, do not appear to bias us towards galaxies of a certain {\it
  physical} size thanks to the wide variety of distances within our
sample.  The distributions of physical sizes of our galaxies before
and after we institute our useability criteria are not significantly
different, as confirmed by a K-S test.  Our final sample
  includes some galaxies taken from the NFGS, which were in fact
  originally chosen to be representative of the galaxy population (see
  \S~\ref{sec:nfgs}).  We will use this subset to investigate our
  results in the context of a broadly representative data set.

\begin{figure}
\epsscale{1.2}
\plotone{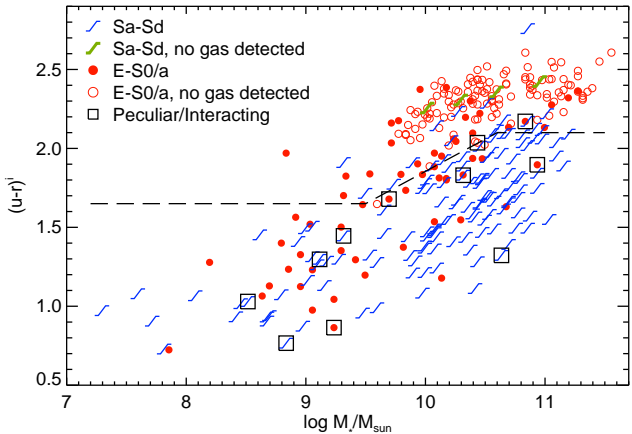}
\caption{Our final NFGS+literature sample plotted in $(u-r)^i$ versus
  stellar mass space.  The dashed line represents the red/blue
  sequence divider, which is based on the analysis of Moffett et
  al.\,(2013, in prep.), which identifies the midpoint between the two
  sequence peaks at each stellar mass.  Our red/blue sequence divider
  is shifted redder than this midpoint by $+$0.1 mag.  The superscript
  $i$ indicates that an internal extinction correction has been
  applied following Moffett et al.}
\label{fig:sample}
\end{figure}

Our final sample, including the quenched and peculiar galaxies, is
shown on the $u-r$ color versus stellar mass plane in
Figure~\ref{fig:sample}.  This sample includes some galaxies with
M$_*<10^{8.5}\,M_{\odot}$, although the blue-centeredness
mass-correction was calibrated with only galaxies above this mass.
These lower mass galaxies do not show unusual behavior within our
results, and are therefore kept in our final sample (see
\S~\ref{sec:mstars2} for further discussion).

 Our final compiled data set is made available in machine
  readable format in the online edition of our paper.  A summary of
  the data is given in Table 3. 

\section{Results}
\label{sec:results}

\begin{figure*}
\plotone{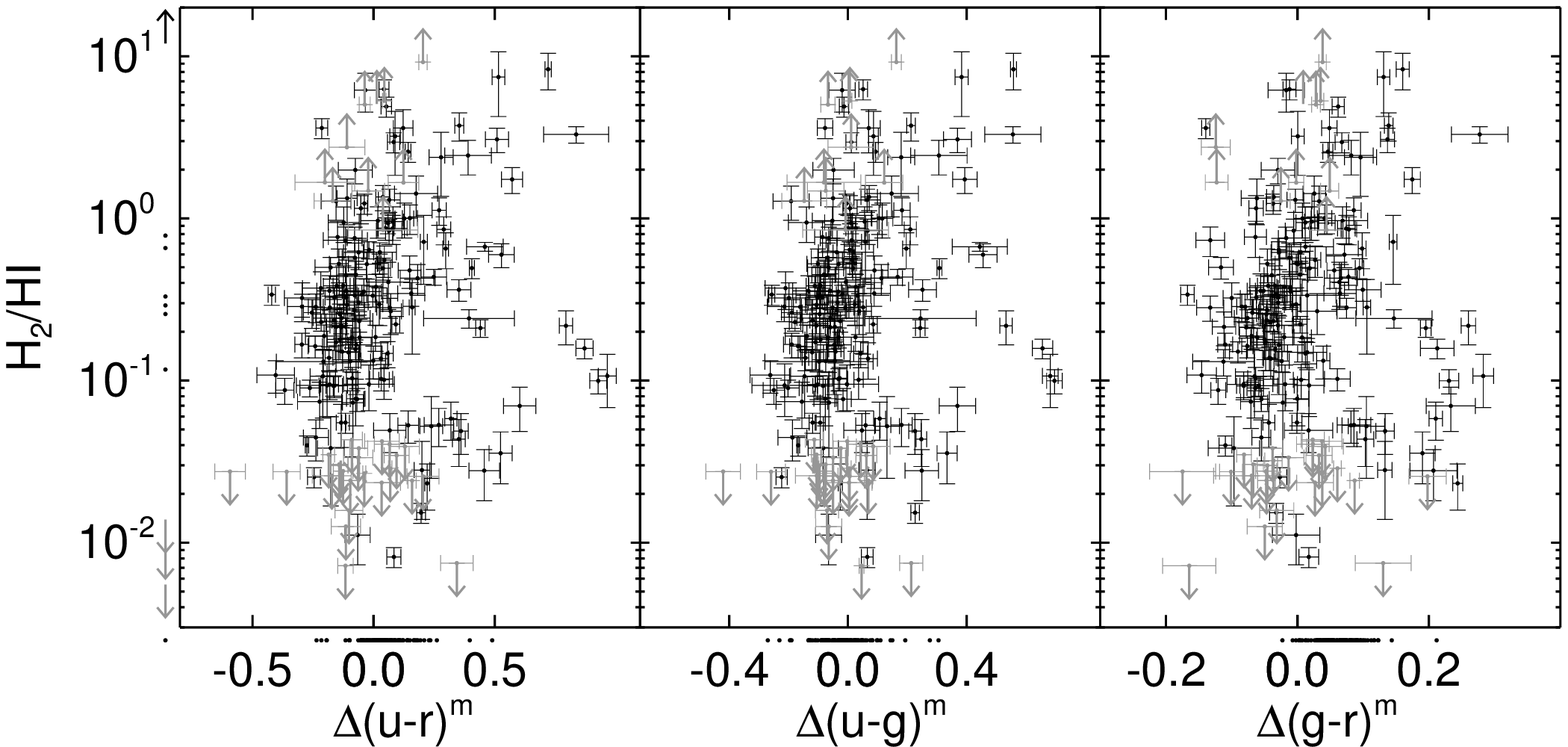}
\caption{Relationship between \htwo/HI and mass-corrected
  blue-centeredness shown using three different color gradients.  Grey
  points denote lower and upper limits in \htwo/HI.  To the left of
  the $y$-axis are dots representing \htwo/HI values for galaxies with
  peculiar morphologies, many clearly interacting.  Due to their
  disturbed state, we cannot measure $\Delta C^m$.  Below each
  $x$-axis are dots showing the measured $\Delta C^m$ for quenched
  galaxies.  All three panels show distinct loci and an empty region
  between them.}
\label{fig:fc}
\end{figure*}

In this section, we describe our findings on the relationship
between global \htwo/HI and mass-corrected blue-centeredness.  These
variables are plotted in Figure~\ref{fig:fc}, which we hereafter refer
to as the ``fueling diagram,'' since it links the fraction of gas
available as direct fuel for star formation and a tracer of recent
interactions that drive the fueling.

In \S~\ref{sec:branches}, we examine the basic structure of the
fueling diagram, which is broken up into three branches.  The left
branch holds most galaxies and shows a positive correlation between
\htwo/HI and $\Delta C^m$, consistent with the idea that molecular gas
content is systematically linked to galaxy interactions as traced by
mass-corrected blue-centeredness.  The right and bottom branches are
well-defined loci that deviate from this expected trend, showing first
increasing then decreasing mass-corrected blue-centeredness as
\htwo/HI decreases.

In \S~\ref{sec:properties}, we explore the properties of galaxies within each branch.
The left branch is largely occupied by a mix of barred and unbarred
spiral galaxies, with a wide range of stellar masses and gas
fractions.  Conversely, the right and bottom branch are almost
completely populated by low mass E/S0 galaxies, specifically
blue-sequence E/S0 galaxies.

Lastly, in \S~\ref{sec:evolution} we use differences in optical color
gradients to explore the evolution of galaxies within the fueling
diagram.  We find no preferred direction of evolution along the left
branch, while the galaxies on the right/bottom branches appear to be
evolving in a clockwise fashion back towards the left branch.
Intriguingly, along this path there are systematic increases in total
gas content and a transition from E/S0 to spiral morphology, which may
signal a major transformation as these galaxies progress back towards
the left branch on the fueling diagram.  We collect and interpret
these findings in \S~\ref{sec:discussion}.

\begin{figure}
\epsscale{1.2}
\plotone{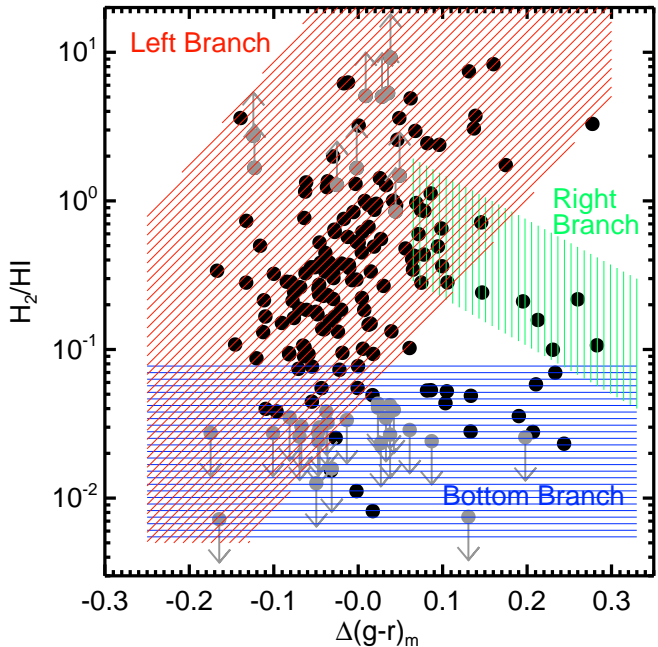}
\caption{Empirically defined branches in \htwo/HI vs. $\Delta C^m$ space.  These data points are the same as in Figure 5c.}
\label{fig:fcregions}
\end{figure}

\subsection{The Distribution of Galaxies in the Fueling Diagram}
\label{sec:distribution}

\subsubsection{The Three Branches}
\label{sec:branches}

The fueling diagram is shown in Figure~\ref{fig:fc}, plotted using
mass-corrected blue-centeredness based on $u-r$, $u-g$, and $g-r$
colors, all of which recreate the same basic structure.  The regions
of parameter space roughly defining the left, right, and bottom
branches are overlaid on the $\Delta(g-r)^m$ data in
Figure~\ref{fig:fcregions}.    

The majority of our sample falls on the left branch, which shows a
positive correlation between \htwo/HI and mass-corrected
blue-centeredness.  There is considerable scatter which appears to
increase above \htwo/HI$\sim$0.5, predominantly biasing galaxies
towards lower $\Delta C^m$.  We attribute at least some of this
scatter to centrally concentrated dust.  Visual inspection has shown
several of the galaxies that scatter to the left of the main trend
have distinct dust features (e.g., M82, which lies at $\Delta
(g-r)^m=-0.14$ and \htwo/HI=3.6).  Galaxies in the ``dusty zone'' do
not have preferentially high inclination (\S~\ref{sec:extinction}).
Rather, their centrally concentrated dust may reflect a shared
evolutionary state (see \S~\ref{sec:mergeleft}).  The right branch of
the fueling diagram, which begins to appear at
$\Delta(g-r)^m\sim0.05-0.1$, is far less populated but still well
defined.  It shows a relationship opposite to the left branch, where
mass-corrected blue-centeredness increases while \htwo/HI decreases.
The bottom branch, which encompasses galaxies below
\htwo/HI$\sim$0.06, shows no clear correlation between \htwo/HI and
mass-corrected blue-centeredness.

Among the galaxies in our sample with gas detections and reliable
$\Delta C^m$ measurements, $\sim$75\% are on the left branch,
$\sim$5\% are on the right branch, and $\sim$20\% are on the bottom
branch.  However, these percentages are very crude since the branches
connect with each other and their definitions are not exact.  We also
stress that our sample is {\it not} statistically representative of
the galaxy population, largely because of the lack of CO
measurements for low-mass galaxies.  Thus, the fractions of galaxies
falling on the left, right, and bottom branches in Fig.~\ref{fig:fc}
cannot be used to infer the true frequency with which galaxies fall on
these parts of the fueling diagram.   If instead we limit
  ourselves to the more representative NFGS subsample, the fractions
  of galaxies on the left, right, and bottom branches are roughly
  65\%, 10\%, and 25\%.  But again, this subsample is only
  approximately representative.

Between these three branches exists a region where no data points lie.
This hole is clearest in the version of the fueling diagram using
$\Delta (g-r)^m$ because the right/bottom branches are most evenly
populated.  For this reason, we default to using the $\Delta(g-r)^m$
version of plot to show further results.

\subsubsection{Is the hole real?}
\label{sec:hole}

The structure seen in the fueling diagram -- particularly the
unoccupied region between the three branches -- does not appear to be
artificially created by our sample restrictions.  As described in
\S~\ref{sec:samples}, our full NFGS+literature sample was unrestricted
in terms of stellar mass, gas content, and star formation properties.
Subsequent restrictions applied were mostly related to the useability
of the data.  The only selection criterion we can reasonably relax is
our limit on the level of permitted CO beam-correction used to
estimate flux missed by the telescope beam.  As discussed in
\S~\ref{sec:beamcorr}, we enforce the corrected-to-measured CO flux to
be less than 1.5.  After relaxing the corrected-to-measured CO flux to
be less than 2, 5, and 10, the three branches of the fueling diagram
remain well defined (although with increased scatter) and no new
galaxies appear to invalidate the existence of the empty region
between the branches.

It is interesting to examine whether peculiar or actively interacting
galaxies could potentially fill the empty region.  These galaxies lack
reliable $\Delta C$ measurements, so we do not plot them within the
fueling diagram.  However, we mark their measured \htwo/HI to the left
of the $y$-axis in Figure~\ref{fig:fc}.  Their \htwo/HI values do not
exclusively cluster in the range spanned by the hole, although some
have the proper values to fall within it.  We conclude that if
galaxies like those represented in our sample ever fall in the empty
region, they must do so only briefly during a phase of rapid
evolution.  The specific reasons why galaxies rarely settle in this
region of parameter space are not immediately apparent.  Detailed
modeling of galaxies may be necessary to explain this phenomenon and
will be the focus of future work.

\subsubsection{X$_{\rm CO}$ Effects}
\label{sec:xco}

\begin{figure*}
\epsscale{1}
\plotone{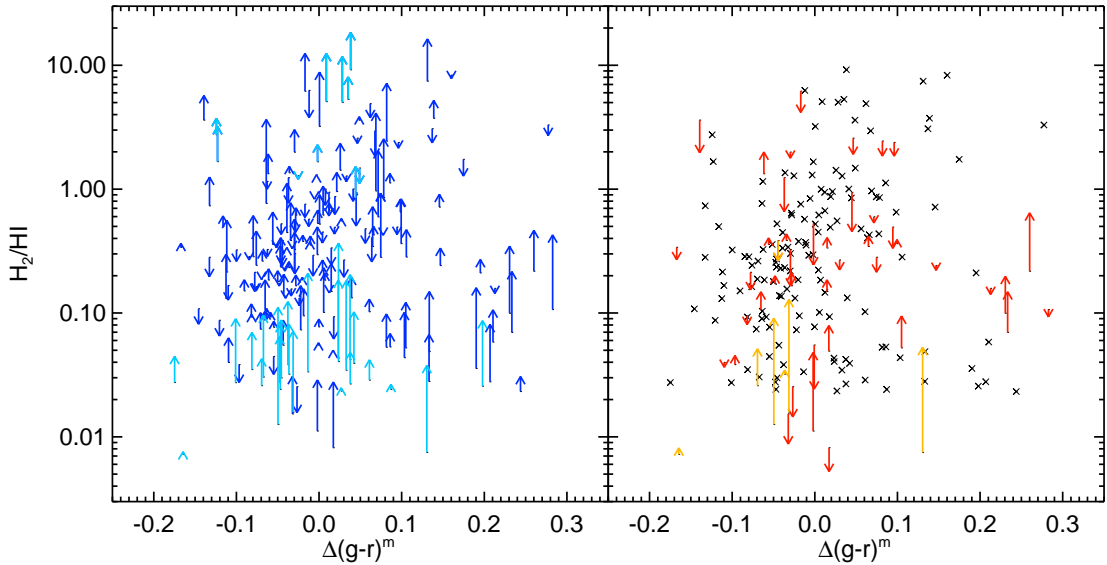}
\caption{The possible effect of variations in X$_{\rm CO}$ on our
  data, with X$_{\rm CO}$ estimated separately for each galaxy using
  $B$-band luminosity (left) and log(O/H) derived from nebular
  emission lines (right).  Lighter arrows denote galaxies with
  \htwo/HI upper or lower limits, and X's denote galaxies where
  log(O/H) is unavailable.  Some upper limits show significant
  adjustments, but X$_{\rm CO}$ variations do not change the overall
  appearance of three distinct branches.}
\label{fig:xco}
\end{figure*}

The structure of galaxies in the fueling diagram appears robust
against possible variations of X$_{\rm CO}$ due to its dependence on
metallicity.  To test this, we use O/H to estimate X$_{\rm CO}$ for
each galaxy, employing the calibration from \citet{Obreschkow09}.
Estimates of O/H from optical line ratios are only available for our
NFGS \citep{Kewley05} and SINGS \citep{Moustakas10} samples and the
studies of \citet{Barone00} and \citet{Taylor98}.  To be able to carry
out this analysis for our full sample, we also estimate X$_{\rm CO}$
for each galaxy from $B$ band luminosity, again using the calibration
of \citet{Obreschkow09}, who find it to be the next most reliable
estimator of X$_{\rm CO}$ after metallicity.  The values of
12+$\log{O/H}$ in our sample (where known) range from 7.9--9.2,
yielding X$_{\rm CO}$ estimates between 0.4 and 10 times the Milky Way
value.  The calibration using $B$-band luminosity yields a similar
range of X$_{\rm CO}$.

Figure~\ref{fig:xco} shows the effect that variable X$_{\rm CO}$ due
to metallicity has on the fueling diagram.  Even with the variation in
X$_{\rm CO}$, the right/bottom branches remain distinct from the left
branch and the hole remains intact, although some of the upper limits
on the bottom branch show significant increases in \htwo/HI due to
their estimates of X$_{\rm CO}$ being several times the Milky Way
value.  These galaxies are mainly dwarfs with stellar masses
$\lesssim$ a few $\times 10^8M_{\odot}$, where this sort of deviation
might be expected.  In general however, the structure of the fueling
diagram remains unchanged.

The above analysis ignores other physics that may alter X$_{\rm CO}$.
Most relevant to our study is the possible {\it decrease} in X$_{\rm
  CO}$ in very high gas surface density regimes where the the ISM
turns almost entirely molecular.  Such a situation could occur as the
result of inflows that drive large amounts of gas to the centers of
galaxies, and has been observed in many systems from ULIRGs to
ordinary spirals \citep{Downes93,Regan00}.  The gas in these extremely
high surface density regions is expected to have increased
temperature relative to molecular clouds in less dense environments
\citep{Narayanan12}, which would be coupled with an increase in the
$F_{\rm CO(2-1)}/F_{\rm CO(1-0)}$ ratio.  Where available, we compare
$F_{\rm CO(2-1)}/F_{\rm CO(1-0)}$ (both beam corrected as in
\S~\ref{sec:beamcorr}) with mass-corrected blue-centeredness and
\htwo/HI, but we find no correlations.  Any evidence of increased
central gas temperature is likely being washed out in our global CO
measurements.  Even if some of the increase in \htwo/HI with $\Delta
C^m$ along the left branch is the result of overestimated X$_{CO}$,
this is still consistent with our physical interpretation that the
left branch of the fueling diagram is the result of galaxy
interactions and inflows (see \S~\ref{sec:mergeleft}).

For the remainder of this paper, we use molecular gas masses estimated
as described in \S~\ref{sec:mgas}, assuming constant X$_{\rm CO}$.

\subsection{Distribution of Galaxy Properties Within the Fueling Diagram}
\label{sec:properties}

\begin{figure*}
\epsscale{1.}
\plotone{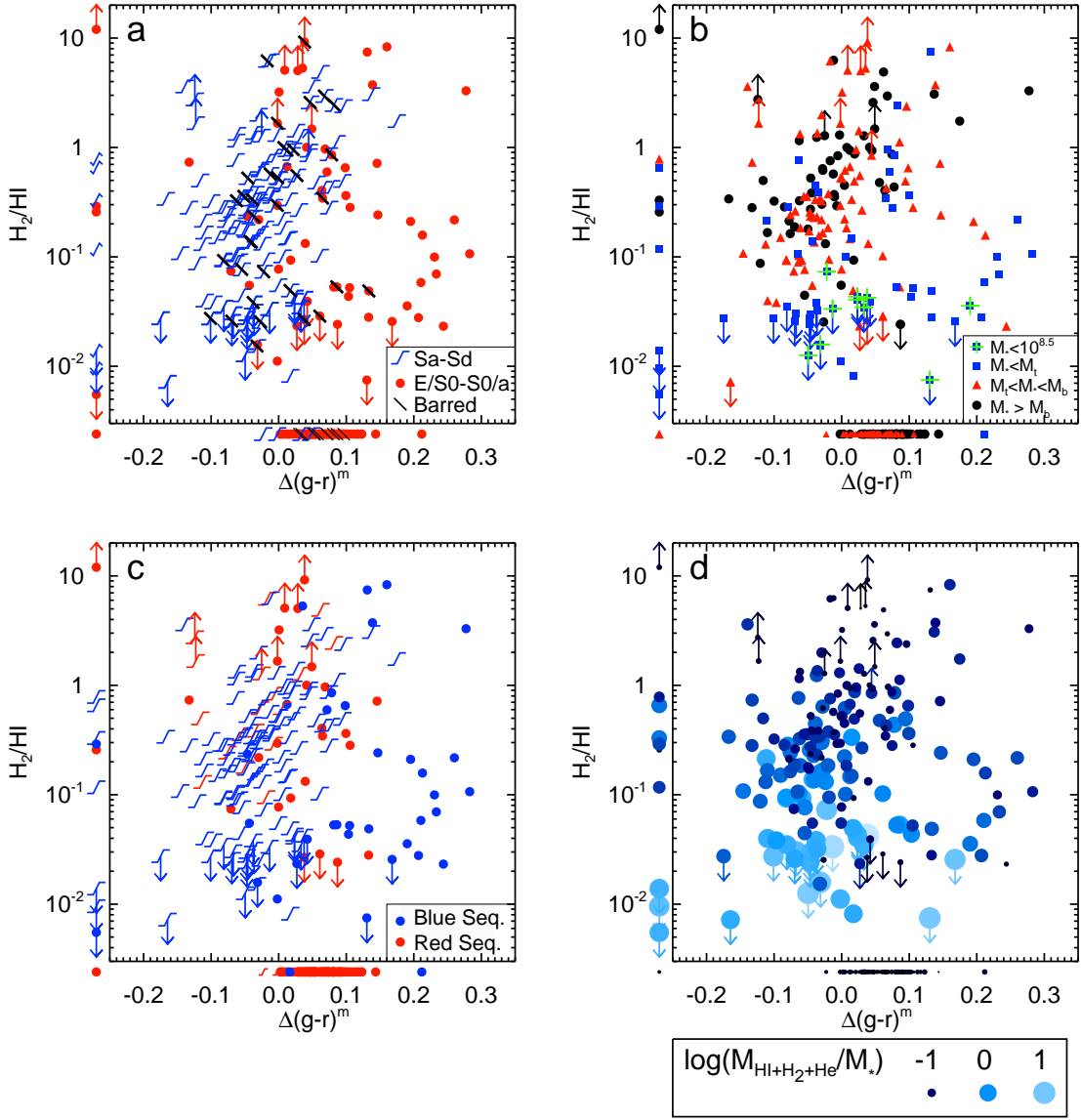}

\caption{({\it a}) Distribution of early (E/S0 - S0a) and late (Sa-Sd)
  morphologies and bars. ({\it b}) Distribution of stellar masses in
  three fundamental mass regimes: stellar mass above $M_b$, between
  $M_t$ and $M_b$, and below $M_t$, where M$_b$ is the bimodality mass
  ($10^{10.5}M_{\odot}$) and $M_t$ is the gas-richness threshold mass
  ($10^{9.7}M_{\odot}$).  Also marked are galaxies that fall below the
  range of our blue-centeredness mass-correction.  See
  \S~\ref{sec:mstars2} for more details on these mass
  regimes. Galaxies with stellar masses outside the mass regime used
  to calibrate the blue-centeredness mass correction are also noted
  ({\it c}) Distribution of red and blue sequence galaxies ({\it d})
  Distribution of $M_{\rm HI+H_2+He}/M_*$.}
\label{fig:3rdvariables}
\end{figure*}

Having established the basic structure of the fueling diagram and its
reliability, we now explore how galaxy properties -- specifically
morphology, the presence of a bar, stellar mass, blue versus red
sequence, and gas content -- distribute themselves throughout the
fueling diagram.  Doing so may provide an understanding of the
physical processes that drive the observed trends.  The galaxy
properties discussed in this section are overlaid on the fueling
diagram in Figure~\ref{fig:3rdvariables} and briefly summarized in
\S~\ref{sec:propsum}.  

\subsubsection{Morphology}
\label{sec:morph}

Figure~\ref{fig:3rdvariables}a displays galaxy morphologies within the
fueling diagram.  All galaxies are classified by eye using SDSS
$g$-band images, but we check our classifications against previously
published types when available.  Galaxies are separated into two
categories: E/S0s (including S0a) and spirals (Sa--Sd).  Using the
distinction between S0a and Sa as the separation between early- and
late-type galaxies may be somewhat sensitive to classification error,
but this division is useful since the presence or absence of extended
spiral arms represents a basic transition in structure likely strongly
linked to star formation history.

We find a clear bimodality in the distribution of E/S0 and spiral
morphologies.  Spiral galaxies almost exclusively fall on the left
branch, and in fact, a Spearman rank test on all spiral galaxies with
\htwo/HI$>$0.06 (to avoid confusion with the bottom branch) confirms a
correlation between \htwo/HI and $\Delta C^m$, with $\sim$5$\sigma$
confidence (for $u-r$, $u-g$, and $g-r$, the probabilities of the
distributions being random are $5\times10^{-8}$, $8\times10^{-7}$, and
$1\times10^{-7}$ respectively).  E/S0s show a much more diverse
distribution throughout the fueling diagram: some occupy the left
branch with spirals, but also and more noticeably the rest almost
completely define the right branch and much of the bottom branch.
Several of the E/S0s on the right/bottom branches are also centrally
concentrated Blue Compact Dwarf galaxies (BCDs), which are lumped with
traditional E/S0s in our simplified morphological classification
scheme.  Towards the left end of the bottom branch, there is a strong
shift in morphology from E/S0 to spiral.

Barred galaxies are also shown in Figure~\ref{fig:3rdvariables}a.
Except for galaxies that have existing bar classifications from the
NFGS or \citet{Nair10}, bar classifications were done by eye using
SDSS $i$-band imaging, since bars are best identified in bands that
trace the stellar light \citep{Eskridge00}.   Bars are
  noticeably absent from all galaxies above $\Delta(g-r)^m\sim0.15$,
  which includes most of the right branch, as well as sections of the
  left and bottom branches.  Within the left branch alone, a Spearman
  Rank test does not suggest a smooth correlation of bar fraction with
  $\Delta C^m$ or with \htwo/HI.  Although our full sample is not
  statistically representative of the galaxy population, we speculate
  that the same processes that produce extreme blue-centeredness may
  destroy bars, while milder processes associated with evolution along
  the left branch do not.  If we limit our examination of bars to only
  the more representative NFGS subsample, we still see no significant
  trend between $\Delta C^m$ and bar fraction within the left branch.
Bars are important galactic structures to put into the context of our
study, since like galaxy interactions, they are thought to enable
inflows of gas to the centers of galaxies.  We discuss bars and
interpret their role in \S~\ref{sec:bars}.

\subsubsection{Stellar Mass}
\label{sec:mstars2}

Figure~\ref{fig:3rdvariables}b displays the distribution of stellar
masses within the fueling diagram.  Instead of examining the
continuous distribution of stellar masses, we divide the data into
three characteristic mass regimes: (1)~$M_* >
M_b=3\times10^{10}\,M_{\odot}$, where $M_b$ is the bimodality mass, a
stellar mass scale above which the population of galaxies goes from
being typically star-forming with disk-like morphology to typically
non-star-forming with spheroidal morphology \citep{Kauffmann03},
(2)~$M_* < M_t=5\times10^9\,M_{\odot}$, where $M_t$ is the
gas-richness threshold mass, below which there is a significant
increase in gas-dominated galaxies (\citealt{KGB}, K13s), and (3)~$M_t
< M_* < M_b$, where bulged spirals with intermediate mass content are
the norm (K13s).  Galaxies with $M_* < 10^{8.5}\,M_{\odot}$ are also
denoted with an extra ``$+$'' symbol in this figure.  These galaxies
technically fall outside the stellar mass range used to calibrate the
blue-centeredness mass-correction (\S~\ref{sec:bc}), and \citet{KJB}
suggest their possibly irregular patterns of star-formation may make
color gradients hard to interpret in relation to higher mass galaxies.
However, we include them in our plots because they show no unusual
distribution relative to the rest of the sample below M$_t$.  Among
these galaxies, the most extreme blue-centeredness is found for
NGC~3738, with $\Delta (g-r)^m=0.19$.  This galaxy is a centrally
concentrated BCD, making it completely consistent with its neighboring
galaxies in the fueling diagram.

 The different stellar mass regimes display patterns within the
 fueling diagram. While galaxies in all three regimes span the full
 range of \htwo/HI, we find almost no galaxies above $M_b$ on the
 right/bottom branches.  Instead, most of the galaxies on these
 branches fall below $M_t$.  There is also a tendency for galaxies
 below $M_t$ to cluster in the lower-left corner of the fueling
 diagram (many having \htwo/HI upper limits), but elsewhere along the
 rising branch they appear to spread roughly evenly, as do galaxies in
 the higher stellar mass regimes.  We note that with our low number
 statistics, the apparent tendency of galaxies on the upper right
 branch to have stellar masses between M$_t$ and M$_b$ is not
 significant and should not be over-interpreted given that the range
 between M$_t$ and M$_b$ is very narrow, and stellar mass estimation
 involves typical errors of $\sim$0.2 dex.

\begin{figure}
\epsscale{1.2} 
\plotone{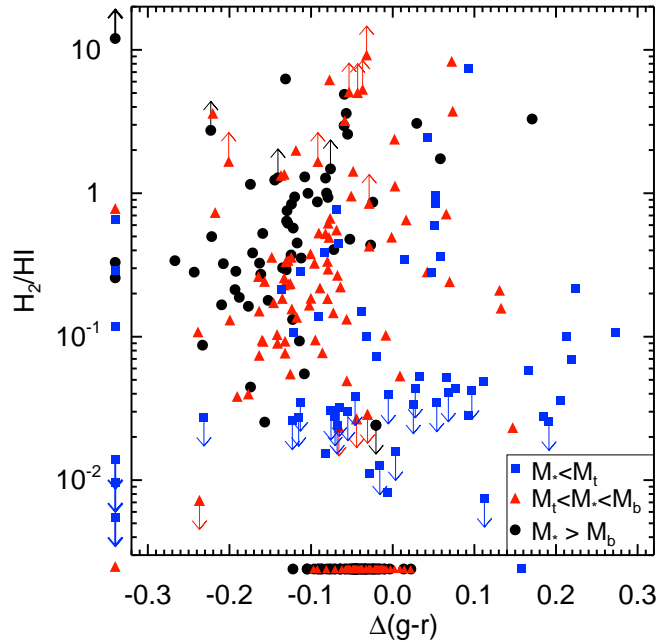}
\caption{The fueling diagram plotted using blue-centeredness without a
  mass correction.  Without the mass correction applied to
  blue-centeredness, there is a tendency for higher stellar mass
  galaxies to have more red-centered color gradients (motivating the
  mass correction), but the underlying structure of the fueling
  diagram remains intact.}
\label{fig:fcnocorr}
\end{figure}

In \S~\ref{sec:bc}, we described our motivation and methods for
subtracting the underlying dependence of blue-centeredness on stellar
mass, yielding mass-corrected blue-centeredness, or $\Delta C^m$.  To
determine how dependent our results are on this mass correction, we
plot the fueling diagram using the uncorrected $\Delta(g-r)$, rather
than $\Delta(g-r)^m$, in Figure~\ref{fig:fcnocorr}.  The general
structure of the fueling diagram remains intact, specifically the
presence of three branches with a hole between them.  The distribution
of different mass regimes illustrates the tendency of high-mass
galaxies to have more red-centered color gradients, which originally
motivated our mass correction.

\subsubsection{Red and Blue Sequences}
\label{sec:sequences}

Figure~\ref{fig:3rdvariables}c shows the distribution of red and
blue-sequence E/S0s and spirals in the fueling diagram, which are
classified based on their positions within the $u-r$ versus stellar
mass plane shown in Figure~\ref{fig:sample}.  The left branch is
composed of a mix of red and blue sequence galaxies.  Intriguingly,
the right and bottom branches are almost completely dominated by blue
sequence galaxies, despite the common assumption that E/S0 galaxies
always fall on the red sequence.  High-mass, red-sequence E/S0
galaxies in our sample are often quenched, i.e., have no detected CO
or HI emission and total gas-to-stellar mass ratio upper limits less
than 0.04M$_{\odot}$.  These systems are shown below the $x$-axis in
Figures~\ref{fig:fc},~\ref{fig:3rdvariables},~\ref{fig:fcnocorr}, and
\ref{fig:agn}.\footnotemark[11]

\footnotetext[11]{We note that the quenched population does not center
  around $\Delta C^m \sim 0$, but rather around
  $\Delta(g-r)^m\sim0.05$ (as one color example).  The $\Delta(g-r)^m$
  values are not due to the presence of excess recent central star
  formation, but rather the fact that the mass correction applied to
  the color gradients was calibrated on star-forming disk galaxies,
  which tend to show more red-centered color gradients than do
  passively evolving galaxies.}

\subsubsection{Gas Content}
\label{sec:gascontent}

Figure~\ref{fig:3rdvariables}d shows the distribution of total
gas-to-stellar mass ratio (total gas = HI+H$_2$ with a 1.4$\times$
mass correction for He) within the fueling diagram.  Broadly speaking,
HI-to-stellar mass ratios and total gas-to-stellar mass ratios on the
left branch are both anti-correlated with \htwo/HI.  However, there is
significant diversity below \htwo/HI$\sim$0.5, where galaxies with
both very high and very low gas fractions fall.  Above
\htwo/HI$\sim$0.5, the gas fractions on the left branch are
consistently low.

Most of the galaxies comprising the right and bottom branches have
substantial gas content, between 10--100\% of their stellar
mass. Along the bottom branch, both HI-to-stellar and total
gas-to-stellar mass ratios appear anti-correlated with mass-corrected
blue-centeredness, that is, gas content is higher on the left, more
red-centered side.  Using a Spearman rank test on all unquenched
bottom branch galaxies (defined as \htwo/HI $<$ 0.06), we find the
 negative correlation of total gas-to-stellar mass ratio
and $\Delta(g-r)^m$ has a  1.6\% chance of being random.  This
probability drops to 0.2\% when restricting the test to
$M_*>10^{8.5}\,M_{\odot}$ (i.e., the mass above which the
blue-centeredness mass correction was calibrated). A few red-sequence
galaxies with very low gas fractions also lie on the bottom branch,
which are among the minority of massive ($M_* > M_t$) E/S0s whose gas
data are not upper limits.

\subsubsection{AGN}
\label{sec:agn} 

\begin{figure}
\epsscale{1.2}
\plotone{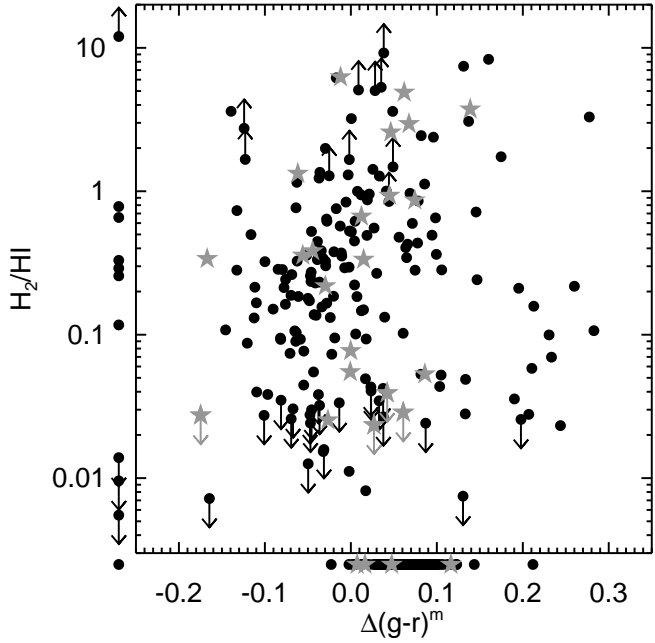}
\caption{Known AGN (stars) within the fueling diagram.  This
  classification is {\it not} complete or uniform in our sample due to
  the limited number of classifications available in the literature
  and incomplete nuclear spectroscopy.}
\label{fig:agn}
\end{figure}

While our sample lacks uniform/complete data for AGN classification,
we display known AGN in Figure~\ref{fig:agn} and briefly discuss them
for two reasons.  First, it is important to note that AGN themselves
are {\it not} the cause of strongly blue-centered color gradients, as
their light contribution is typically too small to have any
significant effect on mass-corrected blue-centeredness, which uses
colors measured over large regions of galaxies.  Second, we note that
AGN are predominantly seen among the high mass galaxies in our sample
(see also Fig.~\ref{fig:3rdvariables}b), with 25 out of 27 AGN hosted
by galaxies with M$_*>{\rm M}_t$.  Their presence, particularly among
the high mass elliptical galaxies which show AGN in the quenched
regime but also among galaxies with detected gas, affects our physical
interpretation of the fueling diagram, and is discussed in
\S~\ref{sec:mergeleft} and \S~\ref{sec:mergebottom}.

\subsubsection{Properties Summary}
\label{sec:propsum}

To summarize, the main properties of each branch of the fueling
diagram are as follows:

The left branch is where spiral galaxies are primarily found.  Many
galaxies here are barred, but without any discernible pattern with
respect to $\Delta C^m$ or \htwo/HI.  Galaxies on the left branch
cover the full range of stellar masses, along with a wide range of HI-
and total gas-to-stellar mass ratios, which broadly behave inversely
to \htwo/HI.

The right and bottom branches are almost entirely populated by
unbarred E/S0 galaxies, although this type distribution transitions
into largely spirals on the left side of the bottom branch, which is
also where we begin to see barred galaxies again.  Stellar masses fall
predominantly below the bimodality mass, and most fall below the threshold
mass.  Gas fractions are moderate to high, between 10\% and 100\% of
the stellar mass, and show a  negative correlation with
  mass-corrected blue-centeredness along the bottom branch.

We also note that the right/bottom branches are dominated by
blue-sequence E/S0 galaxies, i.e., galaxies with spheroidal
morphologies that fall on the blue sequence in color versus stellar
mass space, which are thought to represent a transitional phase
\citep{KGB}.  This fact, as well as the properties of galaxies within
the different branches of the fueling diagram, may reveal important
clues as to how these branches relate to different evolutionary
states, as discussed in \S~\ref{sec:discussion}.

\subsection{Do Galaxies Evolve Within the Fueling Diagram?}
\label{sec:evolution}

If mass-corrected blue-centeredness and \htwo/HI are time-varying
properties, galaxies should evolve within the fueling diagram.  To
determine whether there is any unified direction of evolution within
the fueling diagram or any one of its branches, we search for
systematic offsets between $\Delta (u-g)^m$ and $\Delta (g-r)^m$.
This comparison is useful because $u-g$ color reddens faster than
$g-r$ color after a star forming event, and therefore $\Delta(u-g)^m$
tracks recently enhanced central star formation over shorter time
scales and will return to low values faster than $\Delta(g-r)^m$.
Before comparing these two measurements, we must first account for the
fact that the range of values for $\Delta (u-g)^m$ is larger than for
$\Delta (g-r)^m$ (see Figure~\ref{fig:fc}).   We therefore
  divide $\Delta(u-g)^m$ and $\Delta(g-r)^m$ by their median absolute
  deviations (0.102 and 0.054 respectively, found with the same NFGS
  subsample used to derive Eqs. 4--6) to obtain normalized
  mass-corrected blue-centeredness values, $\Delta(u-g)^{m*}$ and
  $\Delta(g-r)^{m*}$, for which the values cover a similar range while
  not changing the position of zero (i.e., the average gradient for
  the population).
 
\begin{figure}
\epsscale{1.2}
\plotone{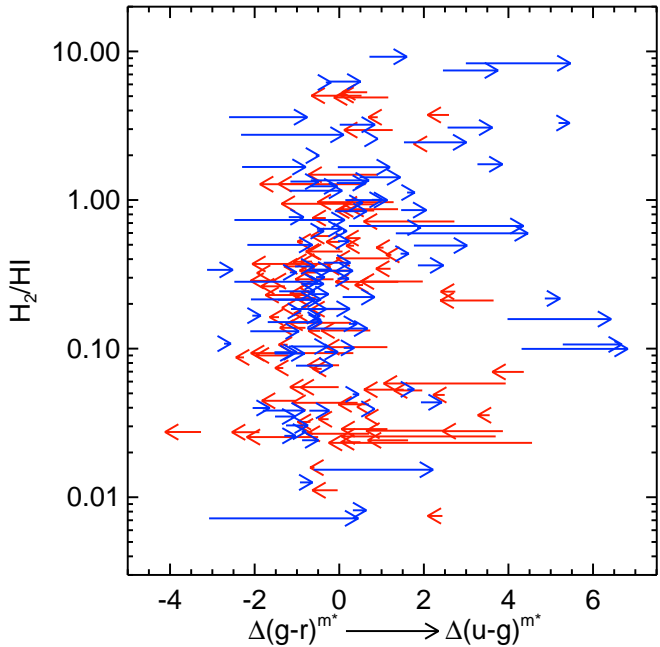}
\caption{Arrows going from $\Delta(g-r)^{m*}$ to $\Delta(u-g)^{m*}$.  Since
  $u-g$ color is more sensitive to high mass short lived stars,
  central enhancements in $u-g$ color should be shorter lived than
  central enhancements in $g-r$ color. The data at the bottom branch
  support this expectation by showing $\Delta(g-r)^{m*}$ more
  blue-centered than $\Delta(u-g)^{m*}$ in most cases, indicating that
  these galaxies' central star formation has ceased and the young
  population is fading.  This color comparison  suggests
  a uniform evolutionary direction of galaxies within this region of
  parameter space. }
\label{fig:evolution}
\end{figure}

In Figure~\ref{fig:evolution}, arrows display the relative positions
of $\Delta(u-g)^{m*}$ (head of arrow) and $\Delta(g-r)^{m*}$
(tail of arrow).  Galaxies on the left branch are directed randomly
left or right along it.  This result suggests that galaxies either do
not evolve throughout this region of the fueling diagram, or the
evolution is not necessarily in a unified direction.  Given the prior
association of mass-corrected blue-centeredness with interactions
\citep{KJB,Kauffmann11}, we suppose that galaxies may oscillate along
the left branch, rising when inflows enhance central star formation
and \htwo/HI (with some of the apparent rise in \htwo/HI possibly
associated with a decrease in X$_{\rm CO}$ due to the central gas
concentration), and fall along the same locus as outer disk gas and
star formation are renewed.

Conversely, the loop composed of the right and bottom branches shows
some partial systematic behavior.   Although the arrows on
  the right branch show no preferred direction, the bottom branch
  shows an excess of galaxies with $\Delta(u-g)^{m*}$ lower than
  $\Delta(g-r)^{m*}$, implying that these galaxies are evolving leftwards
  towards lower mass-corrected blue-centeredness.  Ignoring the
  bottom-left region of the fueling diagram where the left and bottom
  branches cannot be distinguished ($\Delta(g-r)^m < 0.05$), we find
  the frequency of leftward facing arrows on the bottom branch is
  higher than the frequency found on the left branch at the
  4.2$\sigma$ confidence level, assuming the uncertainty in the number
  of rightward/leftward facing arrows follows Poisson statistics.  The
right and bottom branches are likely closely linked (discussed further
in \S~\ref{sec:discussion}), leading to our physical interpretation
that galaxies on the right branch are still undergoing a central
starburst, after which they move to the bottom branch where their
young central stellar population ages and fades.  If star formation is
still progressing progressing on the right branch, then there may be
no reliably predictable difference between $\Delta(u-g)^{m*}$
and $\Delta(g-r)^{m*}$, which would explain the lack of a unified
direction of arrows for right branch galaxies in
Figure~\ref{fig:evolution}.  At this point, we do not have an estimate
of the timescales associated with the evolution along the the right or
bottom branches, but estimating these timescales by comparing the
galaxy colors to stellar population synthesis models will be a focus
of future work.  We also note that the results of
Figure~\ref{fig:evolution} are not dependent on the stellar mass
correction applied to blue-centeredness as the same result is found
even without any mass correction whatsoever.

Since comparing $\Delta (g-r)^{m*}$ and $\Delta (u-g)^{m*}$ gives a clear
direction of evolution along the bottom branch, we can tell that
galaxies here shift from primarily E/S0 to primarily spiral
morphologies, as well as generally toward increasing  gas-to-stellar mass
ratios.  We interpret this pattern as a sign of disk rebuilding in \S~\ref{sec:diskbuilding}.

\section{Discussion}
\label{sec:discussion}

In \S~\ref{sec:results} we presented the fueling diagram relating global \htwo/HI to
recent central star formation enhancements.  Having described the
three-branch distribution of galaxies within the fueling diagram, the
variation of galaxy properties along and between the branches, and the
apparent evolution within the diagram, we now collect these results to
provide an interpretation of the physical processes that drive it.  We
first discuss the role of galaxy interactions in driving \htwo/HI
ratios, considering the importance of the merger mass ratio, stellar
mass, and gas richness of the galaxies involved.  We then explore how
trends within the fueling diagram support a scenario of fresh gas
accretion and stellar disk rebuilding along the bottom branch.  We
finish by reassessing the validity of the assumed link between
mass-corrected blue-centeredness and galaxy interactions in light of
our new results, specifically addressing the role of bars.

\subsection{Mergers and Interactions in the Fueling Diagram}
\label{sec:mergers}

In the following section, we describe the role that mergers play in
driving the evolution in each branch of the fueling diagram.

\subsubsection{The Left Branch and Regions Above It}
\label{sec:mergeleft}

The very existence of the left branch provides support for the idea
that local galaxy interactions play a key role in replenishment of
molecular gas: since mass-corrected blue-centeredness is a signpost of
recent galaxy interactions \citep{KJB}, and it is correlated with
\htwo/HI, galaxy interactions appear to be linked to \htwo/HI in a
systematic way.  The analysis of $\Delta (u-g)^{m*}$ and $\Delta
(g-r)^{m*}$ in \S~\ref{sec:evolution} implies that galaxies on the
left branch may evolve along it in both directions: up as a result of
each galaxy encounter, boosting the galaxy to higher \htwo/HI and
$\Delta C^m$ (with a possible contribution due to overluminous CO),
then down as the molecular reservoir is consumed and the young central
stellar population fades.  Total gas-to-stellar mass ratios support
this scenario.  We expect HI and total gas-to-stellar mass ratios to
decrease at higher $\Delta C^m$ since the gas is being converted into
H$_2$ and then stars, but we also expect a {\it mix} of gas fractions
at low $\Delta C^m$ since galaxies settle here before or after each
burst event while accreting fresh disk gas.  The bulk of the evolution
along the left branch must be driven by minor rather than major
mergers or interactions, since most galaxies appear to retain their
spiral morphologies.  An alternate possibility is that the relation
along the left branch reflects bar-driven inflows.  However, bars do
not show any statistically significant correlation with $\Delta C^m$
along the left branch.  The possible role of bars is explored further
in \S~\ref{sec:bars}

Major mergers in the high stellar mass and/or low gas fraction regime
may help to explain the E/S0s at high \htwo/HI on the left branch (in
contrast, gas-rich, low mass, but comparable mass ratio mergers appear
to follow a different path; see
\S~\ref{sec:mergeright}--\ref{sec:mergebottom}).  Alternatively, these
galaxies could be formed by repeated minor mergers rather than a
single event \citep{Bournaud07}, and they could therefore be galaxies
that have made several oscillations along the left branch.  Either
way, many of the E/S0s at high \htwo/HI on the left branch are likely
moving into the quenched regime, up and off the plot, as their
remaining gas reservoirs have become almost entirely molecular and may
soon be completely depleted.  In addition, some of the high stellar
mass E/S0s at the peak of the rising branch may have previously been
quenched but recently experienced small gas accretion events, possibly
associated with satellite accretion (e.g., \citealt{Martini13}).  Some
of these E/S0s host AGN, as do many of the high mass E/S0s in the
quenched regime.  A gas accretion event could provide fuel for AGN
activity coupled with a central molecular gas concentration, resulting
in a high \htwo/HI ratio with minimal change to $\Delta C^m$ (since
most of the star formation would be occurring very close to the AGN).
Galaxies undergoing this process would jump vertically in the fueling
diagram, between the quenched regime and the peak of the left branch.
This path could place them in the hole of the fueling diagram,
although they would likely move through it relatively quickly.

As previously noted in \S~\ref{sec:branches}, there is substantial
scatter above the left branch towards more red-centered color
gradients, likely caused by internal extinction.  While dust effects
may be altering the measured color gradients, they usefully highlight
galaxies in early stages of star-forming events when the young stars
are still heavily embedded in dust clouds.  We suspect galaxies in
this ``dusty zone'' represent a stage very soon after mergers or
interactions that are mild enough not to have driven the galaxy into
the peculiar morphology region off the plot, shown on the left hand
side of the fueling diagram.  As the star formation progresses, the
dust is likely to clear, allowing these galaxies to develop the
bluer-centered color gradients expected for their \htwo/HI ratios.

\subsubsection{The Right Branch}
\label{sec:mergeright}

We suggest galaxies on the right branch may be the result of gas-rich
mergers, specifically between galaxies of comparable stellar mass.
Along with their high values of mass-corrected blue-centeredness
(suggestive of a strong central starburst), the spheroidal
morphologies of galaxies on the right branch are consistent with
recently violent histories.  Most of these galaxies are classified as
blue-sequence E/S0s, and several are also classified as BCDs (e.g.,
Haro~2, NGC~7077, UM~465).  The existence of blue sequence E/S0s
and BCDs in the same regime of the fueling diagram argues in favor of
them experiencing similar evolutionary processes.  Previous
observational and theoretical studies of BCDs (e.g.,
\citealt{Ostlin01,Pustilnik01,Bekki08}) also support merger driven
evolution.

While most of these galaxies do not show obvious outward signs of a
recent strong interaction in their optical images, such as irregular
structure or tidal features (the lack of these features is actually
built in to our analysis since we do not plot highly
peculiar/interacting systems), smooth optical morphology is not
inconsistent with a merger having recently occurred.  Merger simulations
find the strongest morphological disturbances {\it before} the peak of
induced star formation, which in turn typically occurs before galaxies
land on the right branch, and the complete coalescence of the two
merging galaxy nuclei commonly happens several hundred Myr before the
main starburst event \citep{Lotz08, Lotz10}. The diverse arrow
directions in Figure~\ref{fig:evolution} on the right branch suggest
these galaxies are still actively forming stars well after the merger
remnant has settled.  Signatures of recent mergers may be more obvious
in observations of gas morphology and kinematics.  HI maps can be
extremely useful since they trace extended structure and can retain
signatures of interactions as long as 1 Gyr after the events
\citep{Holwerda11}.  For example, the high resolution HI map of
Haro\,2 \citep{Bravo-Alfaro04} shows the HI kinematic and optical
major axes to be almost perpendicular, consistent with a
merger or recent accretion event.

 Blue-sequence E/S0s are known to emerge primarily below M$_b$, and
 become abundant below M$_t$ \citep{KGB}.  As seen in
 Fig.~\ref{fig:3rdvariables}, the blue E/S0s on the right branch are
 consistent with this pattern.  However, the existence of the right
 branch cannot be driven by stellar mass alone.  Low stellar mass
 galaxies are found throughout the fueling diagram, and if $\Delta
 C^m$ were dictated solely by stellar mass (i.e., equal size bursts
 occurring in higher/low mass galaxies yielding lower/higher $\Delta
 C^m$), then the hole seen in the fueling diagram should not exist.  A
 merger origin for the right branch is more consistent with such a
 large gap between the left and right branches.  Furthermore, gas
 richness likely produces distinct evolutionary tracks for galaxies:
 gas-rich mergers drive galaxies along the right branch, while
 gas-poor mergers drive galaxies into the quenched regime, up and off
 the plot, as discussed in \S~\ref{sec:mergeleft}.  The association of
 the gas-rich merger track with galaxies below M$_b$ and especially
 M$_t$ is a simple consequence of increasing gas richness below those
 scales (\citealt{KGB}, K13s).

\subsubsection{The Bottom Branch} 
\label{sec:mergebottom}

The bottom branch appears to be part of the same evolutionary sequence
as the right branch but at a later stage.  Galaxies here show many of
the same properties as those on the right branch in terms of stellar
mass, gas richness, and prominence of blue-sequence E/S0s.  However,
they show depressed \htwo/HI ratios while uniformly evolving leftwards
in the fueling diagram (\S~\ref{sec:evolution}).  These observations
are all consistent with a scenario where these are post-starburst
galaxies with depleted central gas and fading young central stellar
populations.


\subsection{Evidence for Disk Rebuilding}
\label{sec:diskbuilding}

We have argued that relatively low mass, gas-rich, but roughly
equal-mass-ratio mergers appear to be responsible for the creation of
blue-sequence E/S0 galaxies on the right and bottom branches of the
fueling diagram.  Combining the known direction of evolution on the
bottom branch (\S~\ref{sec:evolution}) with the observed trends in
gas-to-stellar mass ratio and morphology
(\S~\ref{sec:morph},\ref{sec:gascontent}) further implies that these
galaxies may regrow gas and later stellar disks.  For galaxies on the
bottom branch, there is a general {\it increase} in the total
gas-to-stellar mass ratio as $\Delta C^m$ {\it decreases}.  Since
galaxies here appear to be evolving leftward on the fueling diagram,
their total gas content must be growing as their central stellar
populations fade.  In the same direction, there is a transition from
E/S0 morphologies to spiral morphologies.  These combined trends are
consistent with a scenario of fresh outer-disk gas accretion and
eventual conversion into visible spiral arms, and in fact
blue-sequence E/S0s have ideal stellar surface mass densities for
turning gas efficiently into stars, promoting stellar disk rebuilding
(\citealt{KGB}; see also \citealt{Kauffmann06}).  Notwithstanding this
self consistent picture of morphological transformation, there remain
a handful of blue sequence E/S0s in the bottom left corner.  One
possible explanation for their presence is that spiral structure
formation has been inhibited.  Two examples where such inhibition may
be occurring are described in \citet{KGB}: NGC~7360
($\Delta(g-r)^m$=0.027, H$_2$/HI$<$0.023), which hosts
counter-rotating stellar disks, and UGC~9562 ($\Delta(g-r)^m$=-0.002,
H$_2$/HI=0.011), which is a polar ring galaxy.  In both of these
cases, peculiar kinematics may be stifling spiral arm formation.

 Falling below the gas-richness threshold mass M$_t$
  (\S~\ref{sec:mstars2}, K13s) may enable galaxies to evolve leftward
  along the bottom branch.  Bottom branch galaxies typically fall
  below M$_t$ and have not only high gas fractions (as is typical for
  blue-sequence E/S0s in general, \citealt{KGB}, \citealt{Wei10a}),
  but gas fractions that increase as their central starbursts fade.
  The high/increasing gas fractions may indicate rapid gas accretion
  below M$_t$ as argued by K13s.  Theoretical studies of gas accretion
  from the last decade might suggest that M$_t$ reflects the critical
  mass scale for cold-mode accretion \citep{Birnboim03,Keres05}.
  However, \citet{Nelson13} calls this interpretation into question,
  arguing that there is no strong transition from cold to hot mode
  accretion, and that the amount of gas accreted via the cold mode is
  not as large as previously thought.  Regardless of the mode of
  accretion, the high/increasing gas fractions on the bottom branch
  may be explained using the halo mass dependence of gas cooling times
  \citep{Lu11}.  

Conversely, galaxies on the bottom branch that have abnormally low gas
fractions relative to the rest of the population have masses above
M$_t$.  One possible reason may be reduced accretion.  Above the
bimodality mass M$_b$ in particular, both observations and theory
suggest significantly quenched cosmic accretion onto galaxies (e.g.,
\citealt{Gabor12}, K13s).  Another possible reason certain galaxies
might fail to accrete fresh gas on the bottom branch is their
environments.  Galaxies in dense clusters and groups have long been
observed to have depressed HI fractions
\citep{Giovanelli85,Solanes01,Cortese11}, and galaxies near massive
companions are also more likely to be quenched, even in the dwarf
regime below M$_t$ \citep{Geha12}.  One likely example of
neighbor-inhibited accretion within our sample is NGC~3073, a
blue-sequence E/S0 below M$_t$ that has an abnormally low
gas-to-stellar mass ratio, but lies very close to its much larger
companion, NGC~3079.  The fact that NGC~3079 has an AGN and observed
outflows \citep{Cecil01} may also be related to the low gas fraction
in NGC~3073.  Other outliers on the bottom branch, NGC~4111 and
NGC~4270, reside in dense groups.  A full environmental analysis has
not been performed on our sample, but these anecdotal cases hint that
environment, as well as stellar mass, likely determines which galaxies
are capable of regrowing disks.

A final possible explanation of some outliers on the bottom branch may
be that they never even proceeded along the right and bottom branches
to reach their current locations.  As discussed in
\S~\ref{sec:mergeleft}, these may be galaxies from the quenched regime
that have experienced small accretion events, causing them to travel
vertically within the fueling diagram (i.e., through the bottom branch
rather than along it) as they develop/deplete small central molecular
gas concentrations.  In this scenario, such galaxies would only fall
on the bottom branch as an accident of timing.

\subsection{Revisiting the Link Between Mass-Corrected Blue-Centeredness and Galaxy Interactions vs. Bars}
\label{sec:bars}

Throughout this paper, we have made the assumption
that mass-corrected blue-centeredness is linked to galaxy interactions.  This
assumption is motivated by \citet{KJB}, who link blue-centered galaxies
to morphological peculiarities indicative of galaxy interactions.  One
possible issue with this assumption is that E/S0s often lack
morphological peculiarities (partially by their definition of having
smooth light distributions), but may still have blue centers.
Therefore, it is not necessarily obvious that E/S0s have blue centers
for the same reason that clearly disturbed galaxies do.

Our results argue strongly in support of the assumption that high
mass-corrected blue-centeredness implies a recent galaxy encounter,
even in cases where morphological peculiarities are not obvious.  In
fact, the galaxies that have experienced the {\it strongest}
encounters without quenching (gas-rich major mergers of low-mass
galaxies) are probably the blue-sequence E/S0s on the right and bottom
branches that show relatively smooth structure in their optical
images.  A notable exception to the link between blue-centered color
gradients and recent interactions is the existence of the galaxies in
the ``dusty'' zone above the left branch, which appear to be in early,
more dust-embedded stages of star formation.  This implies that there
is a window shortly after the start of induced star formation where
mass-corrected blue-centeredness is a poor indicator of a recent
interaction.

Bars have been suggested as an alternate mechanism for funneling gas
to the centers of galaxies based on direct observations of gas
kinematics \citep{Regan95,Laine99,Regan99}.  The relative importance
of bars versus interactions is not well known due to a lack of large,
homogeneous samples capable of adequately testing both mechanisms.
Recently, \citet{Ellison11} used the abundance of bars and close pairs
in a large sample from the SDSS to argue that bars induce $\sim$3.5
times more central star formation than galaxy interactions, although
the authors note that minor interactions (i.e., pairs with mass ratios
larger than 3/1) are not considered in their analysis.  Whether bars
are a mechanism for inward gas transport completely independent of
galaxy interactions is also unclear.  Galaxy interactions can induce
bar formation \citep{Gerin90,Miwa98}, but at the same time bars can
form in stable disks \citep{Ostriker73,Sellwood81}, and interactions
may actually destroy barred galaxies in some cases
\citep{Berentzen03,Casteels12}.  Studies of bar fractions likewise
give mixed results as to whether bars are related to galaxy
interactions or not \citep{Aguerri09,Li09,Barway11,Lee12}.


Bars do not appear to play any role in evolution along the right and
bottom branches of the fueling diagram.  In fact, they are absent in
these branches, except for the region very near the junction between
the left and bottom branches, which we have already argued is a spiral
rebirth stage long after the inflow event that triggered the
starburst.  This lack of bars in the blue E/S0 population is
consistent with observations of the bar frequency as a function of
morphology \citep{Nair10,Barway11,Lee12}.  One could argue that bars
did induce the original gas inflow and their absence on the right and
bottom branches is caused by the build-up of central mass
concentrations which dissipate bar structures
\citep{Norman96,Shen04,Athanassoula05,Bournaud05}.  However, this
picture would not explain why {\it all} the galaxies are E/S0s.  More
likely, any existing bars were destroyed by violent mergers whose
remnants populate the right and bottom branches of the fueling
diagram.

 Bars are found on the left branch, but the bar fraction
  shows no smooth trend with either $\Delta C^m$ or with \htwo/HI (although
  bars are absent at the top-right of the left branch), and they are
even quite abundant in the bottom-left corner of the fueling diagram
where galaxies show no sign of gas inflow.  This lack of a
smooth correlation may be explained if bar lifetimes are
on the order of a few Gyr \citep{Jogee04,Bournaud05,debattista06}, in
which case bars may drive gas inflow that boosts central gas concentrations
and star formation but also remain well after central gas
concentrations have been depleted and starbursts have ceased
(\citealt{Sheth05,Wang12}, but see \citealt{Ho97} and
\citealt{Sakamoto99} for alternate viewpoints).  This possible
longevity makes interpreting the role bars play on the left branch
difficult.  However, since there are galaxies near the top of the left
branch {\it without} bars, they certainly do not seem to be {\it
  required} to initiate an inflow event.  Bars may increase the
strength of inflows and induce quicker depletion, but at this point it
is unclear whether barred and unbarred galaxies behave
systematically differently on the left branch.


\section{Conclusions}
\label{sec:conclusions}

Using mass-corrected color gradients and \htwo/HI ratios for a sample of
galaxies spanning a broad range of morphologies, stellar masses, and
evolutionary states, we have analyzed the relationship between recent
central star formation enhancements, likely to reflect galaxy
interactions, and global \htwo/HI ratios and total gas content in
galaxies.  We summarize our main results:

\begin{itemize}

\item The parameter space of global \htwo/HI and recently enhanced
  central blueness, which we refer to as the ``fueling diagram,''
  shows a complex relationship composed of three main branches -- the
  left branch, the right branch, and the bottom branch -- with most of
  our galaxies falling on the left branch.  Galaxies in specific
  evolutionary states tend to concentrate in certain regions of the
  diagram (e.g., dusty, early-stage starbursts), or can be represented
  on one axis of the diagram (e.g., quenched systems).  Since our
  sample is not statistically representative of the galaxy population, we
  cannot estimate the frequency with which galaxies fall on
  each branch.

\item The left branch is composed primarily of star-forming spiral
  galaxies with a wide range of stellar masses and gas fractions.  It
  follows a positive correlation between global \htwo/HI and recently
  enhanced central star formation. We interpret this correlation as
  evidence that \htwo/HI ratios are systematically linked to local
  encounters with other galaxies that drive inflows and replenish
  molecular gas reservoirs.  Additionally, apparent enhancement of
  \htwo/HI ratio measurements may be caused by decreased X$_{\rm CO}$
  in the high surface density gas often found in the centers of
  galaxies that experience inflow events.  Galaxies on the left branch
  likely evolve in both directions along it before and after inflow
  events.

\item The right and bottom branches are composed almost exclusively of
  gas-rich blue-sequence E/S0 galaxies with stellar masses below the
  bimodality scale M$_b$ and typically also below the gas-richness threshold
  scale M$_t$.  Several lines of evidence suggest these two branches are
  part of a continuous evolutionary sequence of galaxies formed by
  gas-rich mergers of galaxies with roughly equal masses, which
  results in E/S0 galaxies that are experiencing strong central
  starbursts, depleting their molecular gas, and then fading back
  towards the left branch.

\item The population of galaxies on the bottom branch evolving back
  towards the left branch shows a general increase in total gas
  content and displays a clear transition from primarily E/S0 to
  primarily spiral morphologies.  These results strongly suggest fresh
  cosmic gas accretion and post-merger disk rebuilding in the low mass
  regime.  Our current analysis does not constrain the timescale of
  this regrowth, but this question will be a topic of follow-up
  research.

\item E/S0s above M$_t$ and especially M$_b$ do not obviously move
  along the branches and may instead move vertically in the plot, due
  to minor accretion events associated with nuclear fueling of star
  formation or AGN.

\item Barred galaxies are common on the left branch, although the
  presence of a bar shows no clear correlation with mass-corrected
  blue-centeredness or \htwo/HI.  It is unclear whether bars are
  involved in the small inflow events that drive the evolution on the
  left branch, but bars are likely destroyed in the mergers that
  create the right/bottom branches and therefore play little role in
  these galaxies' evolution.

\end{itemize}

The fueling diagram presented in this study links the amounts of
atomic and molecular gas fuel in a galaxy with a metric for the events
that drive central fueling and HI-to-H$_2$ conversion, providing a
useful framework for understanding how interactions, inflows, and gas
accretion drive the continued growth and evolution of galaxies.  The
movement of galaxies through the interconnected sequences of the
fueling diagram highlights their dynamic evolution.  The left branch
of the fueling diagram holds the ``normal'' star-forming galaxies,
which appear to progress up and down along the left branch during and
after inflow events, with star formation alternately concentrated in
the center vs.\ outer disk.  The right and bottom branches of the
fueling diagram hold the more dramatically transforming galaxies that
have likely experienced recent gas-rich mergers and central
starbursts.  These typically low mass spheroids proceed along the right
and bottom branches until reconnecting with the left branch,
potentially re-forming disk galaxies along the way.  Some galaxies may
proceed through the branches of the fueling diagram multiple times
until quenching mergers drive them off the plot.  Thus the interplay
of bulge building and disk regrowth is a fundamental process revealed
in varying degrees by the distinct evolutionary tracks in the fueling
diagram.

\acknowledgements

We would like to thank Lisa Young for kindly providing the CO(1-0)
spectrum for NGC~5173.  We would also like to thank Amanda Moffett for
useful discussions relating to IRAC reduction and photometry.  We
thank G. Cecil, C. Clemens, F. Heitsch, M. Krumholz, M. Mac Low,
M. Norris, D. Reichart, M. Thornley, and J. Gallimore for useful
discussions and suggestions that helped to improve this research.  We
thank our referee, Erik Rosolowsky, for his helpful comments that
improved this paper.  D. Stark, S. Kannappan, and K. Eckert were
supported in this research by NSF CAREER grant AST-0955368.  D. Stark
and K. Eckert also acknowledge support from GAANN Fellowships and
North Carolina Space Grant Fellowships.  L. Wei was supported in part
by the NSF under the CARMA cooperative agreement and in part by an SMA
Postdoctoral Fellowship.

This work is based on observations carried out with the IRAM 30m
Telescope. IRAM is supported by INSU/CNRS (France), MPG (Germany) and
IGN (Spain).  This work is based on observations carried out with the
Kitt Peak 12 Meter telescope.  The Kitt Peak 12 Meter is operated by
the Arizona Radio Observatory (ARO), Steward Observatory, University
of Arizona.  This work is based in part on observations made with the
{\it Spitzer Space Telescope}, which is operated by the Jet Propulsion
Laboratory, California Institute of Technology under a contract with
NASA.  Funding for SDSS-III has been provided by the Alfred P. Sloan
Foundation, the Participating Institutions, the National Science
Foundation, and the U.S. Department of Energy Office of Science. The
SDSS-III web site is http://www.sdss3.org/.  SDSS-III is managed by
the Astrophysical Research Consortium for the Participating
Institutions of the SDSS-III Collaboration including the University of
Arizona, the Brazilian Participation Group, Brookhaven National
Laboratory, University of Cambridge, Carnegie Mellon University,
University of Florida, the French Participation Group, the German
Participation Group, Harvard University, the Instituto de Astrofisica
de Canarias, the Michigan State/Notre Dame/JINA Participation Group,
Johns Hopkins University, Lawrence Berkeley National Laboratory, Max
Planck Institute for Astrophysics, Max Planck Institute for
Extraterrestrial Physics, New Mexico State University, New York
University, Ohio State University, Pennsylvania State University,
University of Portsmouth, Princeton University, the Spanish
Participation Group, University of Tokyo, University of Utah,
Vanderbilt University, University of Virginia, University of
Washington, and Yale University.  This publication makes use of data
products from the Two Micron All Sky Survey, which is a joint project
of the University of Massachusetts and the Infrared Processing and
Analysis Center/California Institute of Technology, funded by the
National Aeronautics and Space Administration and the National Science
Foundation.  CARMA development and operations are supported by NSF
under a cooperative agreement, and by the CARMA partner universities.
This research has made use of the NASA/IPAC Extragalactic Database
(NED) which is operated by the Jet Propulsion Laboratory, California
Institute of Technology, under contract with the National Aeronautics
and Space Administration.

\clearpage

\clearpage

\begin{turnpage}
\scriptsize
\tablecaption{New IRAM 30m CO(1-0) and CO(2-1) Measurements}
\begin{deluxetable}{lccccc|ccccc|ccc}
\tablewidth{0pt}
\tablehead{
 & \multicolumn{5}{c}{{\bf CO(1-0)}} &  \multicolumn{5}{c}{{\bf CO(2-1)}} & & & \\
\colhead{Name} & \colhead{Flux} & \colhead{cz} & \colhead{W$_{50}$\tablenotemark{   a}} & \colhead{rms} & \colhead{Range\tablenotemark{   b}} & \colhead{Flux} & \colhead{cz} & \colhead{W$_{50}$\tablenotemark{   a}} & \colhead{rms} & \colhead{Range\tablenotemark{   b}} & \colhead{R$_{25}$} & \colhead{Log M$_{\rm H_2}$\tablenotemark{c}} & \colhead{Log M$_{\rm {H_2},{corr}}$\tablenotemark{   c,d}}\\
\colhead{} & \colhead{Jy km s$^{-1}$} & \colhead{km s$^{-1}$} & \colhead{km s$^{-1}$} & \colhead{mJy} & \colhead{km s$^{-1}$} & \colhead{Jy km s$^{-1}$} & \colhead{km s$^{-1}$} & \colhead{km s$^{-1}$} & \colhead{mJy} & \colhead{km s$^{-1}$} & \colhead{\arcsec} & \colhead{M$_{\odot}$} & \colhead{M$_{\odot}$} \\
}
\startdata
UGC439 & 38.81$\pm$1.42 &         5304 & 116$\pm$3 & 32.17 & 5208--5395 & 41.91$\pm$1.32 &         5306 & 112$\pm$3 & 32.06 & 5221--5383 & 34.0 & 9.38$\pm$0.02 & 9.71$\pm$0.02\\
UGC1154 & 19.46$\pm$0.68 &         7686 & 98$\pm$20 & 11.40 & 7596--7937 & 21.46$\pm$1.16 &         7729 & 171$\pm$9 & 20.85 & 7591--7889 & 24.2 & 9.43$\pm$0.02 & 9.62$\pm$0.02\\
UGC1155 & 8.21$\pm$0.63 &         3190 & 127$\pm$9 & 14.15 & 3087--3275 & 11.95$\pm$1.63 &         3185 & [121$\pm$7] & 37.37 & 3113--3295 & 20.1 & 8.26$\pm$0.03 & 8.39$\pm$0.03\\
NGC2780 & 16.96$\pm$0.92 &         1992 & 203$\pm$6 & 17.36 & 1863--2132 & 20.89$\pm$1.53 &         1973 & 79$\pm$19 & 33.67 & 1886--2084 & 31.7 & 8.27$\pm$0.02 & 8.51$\pm$0.02\\
NGC2844 & 28.73$\pm$1.70 &         1523 & 223$\pm$22 & 25.75 & 1300--1720 & 29.51$\pm$1.83 &         1492 & [266$\pm$14] & 30.92 & 1321--1658 & 54.9 & 8.31$\pm$0.03 & 8.69$\pm$0.03\\
NGC3011 & $<$1.57 & \nodata & \nodata & 12.76 & 1436--1598 & $<$4.63 & \nodata & \nodata & 37.60 & 1436--1598 & 24.8 & $<$7.06 & $<$7.13\\
IC2520 & 65.96$\pm$0.88 &         1259 & 138$\pm$4 & 15.44 & 1122--1437 & 100.50$\pm$1.28 &         1260 & 108$\pm$2 & 24.57 & 1136--1399 & 26.6 & 8.51$\pm$0.01 & 8.73$\pm$0.01\\
UGC5378 & 17.09$\pm$0.88 &         4151 & 188$\pm$13 & 15.32 & 3993--4310 & 26.49$\pm$0.95 &         4165 & 129$\pm$8 & 18.89 & 4022--4264 & 38.2 & 8.85$\pm$0.02 & 9.09$\pm$0.02\\
NGC3213 & 8.62$\pm$0.69 &         1377 & 113$\pm$6 & 16.18 & 1291--1464 & 8.26$\pm$1.92 &         1366 & [83$\pm$41] & 32.77 & 1231--1560 & 36.5 & 7.74$\pm$0.03 & 8.03$\pm$0.03\\
IC2591 & 5.34$\pm$1.25 &         6706 & [246$\pm$36] & 18.92 & 6479--6899 & \nodata & \nodata & \nodata & \nodata & \nodata & 39.2 & 8.76$\pm$0.10 & 9.04$\pm$0.10\\
UGC6003 & 8.10$\pm$0.75 &         5752 & 83$\pm$8 & 15.24 & 5630--5860 & \nodata & \nodata & \nodata & \nodata & \nodata & 19.1 & 8.80$\pm$0.04 & 8.84$\pm$0.04\\
UGC6104 & $<$2.96 & \nodata & \nodata & 18.88 & 2805--3067 & $<$4.91 & \nodata & \nodata & 31.40 & 2805--3067 & 50.9 & $<$7.83 & $<$8.11\\
NGC3633 & 81.45$\pm$1.49 &         2604 & 307$\pm$4 & 22.93 & 2406--2813 & 147.24$\pm$3.30 &         2603 & 271$\pm$11 & 49.66 & 2406--2831 & 41.6 & 9.15$\pm$0.01 & 9.39$\pm$0.01\\
IC692 & 1.83$\pm$0.30 &         1192 & 62$\pm$9 & 12.07 & 1161--1222 & $<$2.36 & \nodata & \nodata & 24.07 & 1106--1208 & 24.2 & $<$6.97 & $<$7.02\\
UGC6545 & 14.38$\pm$0.73 &         2657 & 94$\pm$18 & 12.77 & 2504--2816 & 24.05$\pm$0.73 &         2641 & 105$\pm$8 & 14.13 & 2534--2793 & 46.3 & 8.45$\pm$0.02 & 8.72$\pm$0.02\\
UGC6570 & 14.07$\pm$0.58 &         1621 & 96$\pm$7 & 12.77 & 1505--1701 & 24.03$\pm$0.56 &         1621 & 96$\pm$3 & 13.74 & 1539--1696 & 35.4 & 8.10$\pm$0.02 & 8.19$\pm$0.02\\
UGC6637 & 2.56$\pm$0.51 &         1875 & [41$\pm$45] & 10.35 & 1732--1964 & 4.00$\pm$0.38 &         1866 & [100$\pm$6] & 10.05 & 1795--1936 & 27.0 & 7.45$\pm$0.09 & 7.50$\pm$0.09\\
UGC6805 & 8.09$\pm$0.77 &         1159 & 121$\pm$7 & 14.07 & 1050--1338 & 8.34$\pm$1.23 &         1175 & [49$\pm$18] & 29.70 & 1059--1224 & 20.3 & 7.56$\pm$0.04 & 7.60$\pm$0.04\\
IC746 & 5.84$\pm$0.88 &         4978 & [31$\pm$29] & 17.99 & 4861--5091 & 4.46$\pm$1.32 &         4970 & [75$\pm$6] & 40.81 & 4924--5025 & 43.7 & 8.56$\pm$0.07 & 8.81$\pm$0.07\\
UGC7020A & 24.82$\pm$0.66 &         1519 & 110$\pm$4 & 13.60 & 1394--1618 & 39.23$\pm$1.10 &         1520 & 107$\pm$4 & 22.13 & 1412--1651 & 33.8 & 8.29$\pm$0.01 & 8.37$\pm$0.01\\
UGC7129 & 22.69$\pm$0.91 &          959 & 80$\pm$4 & 22.47 & 886--1043 & 25.71$\pm$0.91 &          959 & 55$\pm$4 & 25.20 & 896--1021 & 41.9 & 7.92$\pm$0.02 & 8.24$\pm$0.02\\
NGC5173 & 8.98$\pm$1.22 &         2438 & [154$\pm$2] & 25.00 & 2302--2531 & \nodata & \nodata & \nodata & \nodata & \nodata & 35.0 & 8.23$\pm$0.06 & 8.34$\pm$0.06\\
NGC5338 & 13.31$\pm$0.89 &          845 & 64$\pm$4 & 19.92 & 752--943 & 29.16$\pm$1.62 &          849 & 59$\pm$5 & 36.53 & 725--915 & 56.1 & 7.19$\pm$0.03 & 7.38$\pm$0.03\\
NGC5596 & $<$2.15 & \nodata & \nodata & 15.03 & 3058--3276 & $<$3.11 & \nodata & \nodata & 21.76 & 3058--3276 & 30.8 & $<$7.79 & $<$7.87\\
NGC5762 & $<$2.61 & \nodata & \nodata & 19.05 & 1688--1888 & $<$2.99 & \nodata & \nodata & 21.86 & 1688--1888 & 44.2 & $<$7.44 & $<$7.82\\
UGC9562 & 3.54$\pm$0.95 &         1256 & [224$\pm$17] & 19.61 & 1144--1368 & 8.62$\pm$2.13 &         1330 & [95$\pm$47] & 39.18 & 1105--1388 & 27.5 & 7.39$\pm$0.12 & 7.45$\pm$0.12\\
UGC9562NE\tablenotemark{   e} & $<$1.09 & \nodata & \nodata & 20.61 & 1262--1292 & 1.48$\pm$0.07 &  1272
 & 
[8$\pm$4]
 & 47.12 & 1262--1292 & \nodata & \nodata & \nodata\\
UGC9562SW\tablenotemark{   f} & $<$0.78 & \nodata & \nodata & 16.18 & 1035--1060 & 1.22$\pm$0.13 &  1049
 & 
[8$\pm$2]
 & 39.59 & 1035--1060 & \nodata & \nodata & \nodata\\
IC1066 & 11.18$\pm$0.87 &         1572 & 107$\pm$14 & 18.32 & 1488--1704 & 9.43$\pm$0.83 &         1566 & [85$\pm$14] & 21.18 & 1505--1652 & 46.0 & 7.99$\pm$0.03 & 8.33$\pm$0.03\\
NGC5874 & 10.72$\pm$0.59 &         3148 & 204$\pm$9 & 10.60 & 3011--3306 & 9.77$\pm$1.21 &         3132 & [133$\pm$20] & 18.94 & 2990--3383 & 64.4 & 8.48$\pm$0.02 & 9.00$\pm$0.02\\
NGC7328 & 48.65$\pm$1.71 &         2793 & 202$\pm$17 & 28.06 & 2656--3014 & 46.68$\pm$1.50 &         2803 & 158$\pm$16 & 25.84 & 2675--3000 & 62.5 & 8.98$\pm$0.02 & 9.36$\pm$0.02\\
NGC7360 & $<$2.29 & \nodata & \nodata & 12.45 & 4517--4877 & $<$3.79 & \nodata & \nodata & 20.64 & 4517--4877 & 36.5 & $<$8.07 & $<$8.15\\
UGC12265N & 13.29$\pm$1.10 &         5651 & [246$\pm$27] & 17.15 & 5435--5832 & \nodata & \nodata & \nodata & \nodata & \nodata & 16.8 & 9.00$\pm$0.04 & 9.03$\pm$0.04\\
NGC7460 & 40.41$\pm$0.81 &         3183 & 156$\pm$4 & 16.18 & 3052--3295 & 41.74$\pm$1.11 &         3177 & 137$\pm$4 & 21.22 & 3054--3315 & 44.8 & 9.01$\pm$0.01 & 9.43$\pm$0.01\\
NGC7537 & 35.37$\pm$1.93 &         2643 & 210$\pm$20 & 32.05 & 2500--2847 & 40.80$\pm$2.18 &         2657 & [218$\pm$7] & 39.36 & 2538--2834 & 68.3 & 8.77$\pm$0.02 & 9.14$\pm$0.02\\
\enddata
\tablecomments{$^{\rm{a}}$ Brackets denote galaxies with S/N$<$6 where linewidths become extremely unreliable and should be used with caution.  $^{\rm{b}}$ Range of velocities used in integration.  $^{\rm{c}}$ Masses include factor of 1.4 to account for Helium.  $^{\rm{d}}$ Beam-corrected H$_2$ mass (see \S~\ref{sec:beamcorr}).  $^{\rm{e}}$ Offset from center of UGC9562 by $+9\arcsec$,$+11.5\arcsec$ to observe polar ring.  Upper limit integration on CO(1-0) flux based on integration range for CO(2-1) detection.  CO(2-1) measurements done at a resolution of 2.6 km\,s$^{-1}$ due to extremely small linewidth.  $^{\rm{f}}$ Offset from center of UGC9562 by $-6\arcsec$,$-9.7\arcsec$ to observe polar ring.  Upper limit integration on CO(1-0) flux based on integration range for CO(2-1) detection.  CO(2-1) measurements done at a resolution of 2.6 km\,s$^{-1}$ due to extremely small linewidth.  }
\end{deluxetable}
\end{turnpage}

\clearpage

\begin{deluxetable}{lccccccccc}
\scriptsize
\tablecaption{New ARO 12m CO(1-0) Measurements}
\tablewidth{0pt}
\tablehead{
\colhead{Name} & \colhead{Flux} & \colhead{cz} & \colhead{W$_{50}$\tablenotemark{   a}} & \colhead{rms} & \colhead{Range\tablenotemark{   b}} & \colhead{R$_{25}$} & \colhead{Log M$_{\rm H_2}$\tablenotemark{c}} & \colhead{Log M$_{\rm {H_2},{corr}}$\tablenotemark{   c,d}}\\
\colhead{} & \colhead{Jy km s$^{-1}$} & \colhead{km s$^{-1}$} & \colhead{km s$^{-1}$} & \colhead{mJy} & \colhead{km s$^{-1}$} & \colhead{\arcsec} & \colhead{M$_{\odot}$} & \colhead{M$_{\odot}$} \\
}
\startdata
UGC439 & 32.37$\pm$2.49 &         5301 & 129$\pm$5 & 60.80 & 5218--5379 & 34.0 & 9.30$\pm$0.03 & 9.39$\pm$0.03\\
UGC1154 & 45.52$\pm$4.53 &         7770 & [194$\pm$7] & 81.05 & 7632--7932 & 24.2 & 9.79$\pm$0.04 & 9.84$\pm$0.04\\
UGC1155 & $<$22.34 & \nodata & \nodata & 158.21 & 3052--3264 & 20.1 & $<$8.69 & $<$8.72\\
NGC2780 & 40.38$\pm$4.23 &         1976 & [204$\pm$5] & 70.20 & 1760--2109 & 31.7 & 8.64$\pm$0.05 & 8.70$\pm$0.05\\
UGC4902* & 5.79$\pm$2.49 & 1633 & [67$\pm$18] & 93.91 & 1600--1667 & 40.2 & 7.83$\pm$0.18 & 7.85$\pm$0.18\\
NGC3032* & 114.11$\pm$10.79 &  1555
 & 
[151$\pm$5]
 & 258.50 & 
1474--1642
 & 60.0 & 8.90$\pm$0.03 & 9.02$\pm$0.03\\
IC2520 & 105.03$\pm$6.92 &         1219 & 147$\pm$19 & 124.77 & 1072--1368 & 26.6 & 8.71$\pm$0.03 & 8.76$\pm$0.03\\
UGC5744 & 41.90$\pm$3.64 &         3366 & [80$\pm$18] & 88.97 & 3270--3431 & 22.8 & 9.10$\pm$0.04 & 9.14$\pm$0.04\\
NGC3419* & 53.66$\pm$4.91 &  3040
 & 
[107$\pm$34]
 & 91.01 & 
2884--3164
 & 36.9 & 9.05$\pm$0.04 & 9.07$\pm$0.04\\
UGC6003 & 9.40$\pm$1.96 &         5833 & [184$\pm$12] & 44.79 & 5741--5925 & 19.1 & 8.87$\pm$0.09 & 8.87$\pm$0.09\\
NGC3633 & 147.45$\pm$7.49 &         2588 & 325$\pm$9 & 119.83 & 2401--2776 & 41.6 & 9.41$\pm$0.02 & 9.47$\pm$0.02\\
UGC6545 & 33.21$\pm$3.95 &         2670 & [30$\pm$26] & 104.81 & 2564--2701 & 46.3 & 8.81$\pm$0.05 & 8.89$\pm$0.05\\
UGC6570 & 17.86$\pm$3.02 &         1574 & [106$\pm$7] & 77.30 & 1504--1651 & 35.4 & 8.21$\pm$0.07 & 8.22$\pm$0.07\\
NGC3773* & 9.14$\pm$2.33 &  1002
 & 
[82$\pm$9]
 & 79.92 & 
961--1043
 & 35.1 & 7.05$\pm$0.11 & 7.07$\pm$0.11\\
NGC3870* & 7.96$\pm$1.68 &  747
 & 
[69$\pm$13]
 & 62.56 & 
712--782
 & 31.2 & 7.27$\pm$0.09 & 7.34$\pm$0.09\\
UGC6805 & 13.79$\pm$3.16 &         1096 & [32$\pm$33] & 85.13 & 1071--1204 & 20.3 & 7.79$\pm$0.10 & 7.80$\pm$0.10\\
UGC7020A & 35.23$\pm$4.25 &         1536 & [138$\pm$11] & 93.26 & 1450--1650 & 33.8 & 8.44$\pm$0.05 & 8.46$\pm$0.05\\
UGC7129 & 72.99$\pm$7.12 &          930 & [91$\pm$24] & 154.42 & 822--1027 & 41.9 & 8.43$\pm$0.04 & 8.52$\pm$0.04\\
NGC5173 & $<$20.31 & \nodata & \nodata & 152.83 & 2373--2561 & 35.0 & $<$8.58 & $<$8.60\\
NGC5338 & 23.56$\pm$3.46 &          807 & [40$\pm$29] & 89.07 & 751--896 & 56.1 & 7.44$\pm$0.06 & 7.48$\pm$0.06\\
NGC7077 & $<$14.56 & \nodata & \nodata & 142.54 & 1086--1198 & 25.9 & $<$7.76 & $<$7.77\\
NGC7328 & 98.88$\pm$8.56 &         2763 & [148$\pm$55] & 134.02 & 2592--2984 & 62.5 & 9.29$\pm$0.04 & 9.41$\pm$0.04\\
NGC7328E\tablenotemark{e} & 37.41$\pm$6.42 &  2930
 & 
[99$\pm$8]
 & 154.74 & 
2843--3008
 & \nodata & \nodata & \nodata\\
NGC7328W\tablenotemark{f} & 47.30$\pm$6.12 &  2763
 & 
[217$\pm$9]
 & 112.62 & 
2641--2924
 & \nodata & \nodata & \nodata\\
UGC12265N & 21.33$\pm$2.95 &         5707 & [212$\pm$8] & 53.57 & 5570--5862 & 16.8 & 9.21$\pm$0.06 & 9.21$\pm$0.06\\
NGC7460 & 61.89$\pm$6.70 &         3185 & [163$\pm$11] & 136.14 & 3080--3313 & 44.8 & 9.19$\pm$0.05 & 9.32$\pm$0.05\\
NGC7537 & 51.20$\pm$6.51 &         2622 & [135$\pm$51] & 118.60 & 2543--2833 & 68.3 & 8.93$\pm$0.06 & 9.06$\pm$0.06\\
\enddata
\tablecomments{$^{\rm{a}}$ Brackets denote galaxies with S/N$<$6 where linewidths become extremely unreliable and should be used with caution.  $^{\rm{b}}$ Range of velocities used in integration.  $^{\rm{c}}$ Mass includes factor of 1.4 to account for Helium.  $^{\rm{d}}$ Beam-corrected H$_2$ mass (see \S~\ref{sec:beamcorr}).  $^{\rm e}$  Offset from center of NGC7328 by $+28.5\arcsec$,$+1.2\arcsec$.  $^{\rm f}$ Offset from center of NGC7328 by $-27\arcsec$,$-1.9\arcsec$.  $^*$ Non-NFGS galaxy.  }
\end{deluxetable}

\clearpage
\scriptsize
\tablecaption{Fueling Diagram Catalog Description}
\begin{deluxetable}{ll}
\tablewidth{0pt}
\tablehead{\colhead{Column} & \colhead{Description}}
\startdata
1 & Object name	\\					       
2 & Right Ascension		\\					       
3 & Declination		\\					       
4 & Assumed distance to galaxy	\\					       
5 & Inclination	(see \S2.3.1)	\\					       
6 & $u-r$ color  	\\						       
7 & Stellar mass	\\						       
8 & Uncertainty in stellar mass	\\				       
9 & Reference for HI mass 	\\					       
10 & Reference for uncorrected H$_2$ mass	\\			       
11 & H$_2$ upper limit flag	\\					       
12 & H$_2$ mass after beam correction 	\\		       
13 & Uncertainty in H$_2$ mass after beam correction	\\		       
14 & X$_{\rm{CO}}$ derived from $B$ band luminosity	\\		       
15 & X$_{\rm{CO}}$ derived from O/H 		\\			       
16 & $u-r$ blue-centeredness	\\					       
17 & Uncertainty in $u-r$ blue-centeredness	\\			       
18 & Stellar mass-corrected $u-r$ blue-centeredness	\\		       
19 & Uncertainty in stellar mass-corrected $u-r$ blue-centeredness     	\\ 
20 & $u-g$ blue-centeredness	\\					       
21 & Uncertainty in $u-g$ blue-centeredness	\\			       
22 & Stellar mass-corrected $u-g$ blue-centeredness	\\		       
23 & Uncertainty in stellar mass-corrected $u-g$ blue-centeredness 	\\     
24 & $g-r$ blue-centeredness	\\					       
25 & Uncertainty in $g-r$ blue-centeredness		\\		       
26 & Stellar mass-corrected $g-r$ blue-centeredness	\\		       
27 & Uncertainty in stellar mass-corrected $g-r$ blue-centeredness  	\\    
28 & Morphology (see \S3.2.1)   	\\			       
29 & Flag to indicate peculiar galaxy (see \S2.3.7)	\\			
30 & Flag to indicate presence of bar	\\				       
31 & Reference for bar classification    	\\ 	                       
\enddata
\end{deluxetable}

\end{document}